\documentclass[trackchanges]{aastex63}
\usepackage{lineno}

\received{}
\revised{}
\accepted{}
\submitjournal{AJ}

\shorttitle{Densified Pupil Spectrograph as Radial Velocimetry}
\shortauthors{Matsuo et al.}

\begin{document}

\title{Densified pupil spectrograph as high-precision radial velocimetry: From direct measurement of the Universe's expansion history to characterization of nearby habitable planet candidates}

\correspondingauthor{Taro Matsuo}
\email{matsuo@u.phys.nagoya-u.ac.jp}

\author[0000-0001-7694-5885]{Taro Matsuo}
\affiliation{Department of Particle and Astrophysics, Graduate School of Science, Nagoya University \\ Furocho, Chikusa-ku, Nagoya, Aichi 466-8601, Japan}
\affiliation{NASA Ames Research Center, Moffett Field, CA 94035, USA}

\author[0000-0002-8963-8056]{Thomas P. Greene}
\affiliation{NASA Ames Research Center, Space Science and Astrobiology Division, M.S. 245-6, Moffett Field, CA 94035, USA}

\author[0000-0001-7066-1240]{Mahdi Qezlou}
\affiliation{Department of Physics \& Astronomy, University of California Riverside, CA 92521, USA}

\author[0000-0001-5803-5490]{Simeon Bird}
\affiliation{Department of Physics \& Astronomy, University of California Riverside, CA 92521, USA}

\author{Kiyotomo Ichiki}
\affiliation{Department of Particle and Astrophysics, Graduate School of Science, Nagoya University \\ Furocho, Chikusa-ku, Nagoya, Aichi 466-8601, Japan}
\affiliation{Kobayashi-Maskawa Institute for the Origin of Particles and the Universe, Nagoya University, Chikusa-ku, Nagoya 464-8602, Japan}

\author[0000-0002-2786-0786]{Yuka Fujii}
\affiliation{National Astronomical Observatory of Japan, 2-21-1, Osawa, Mitaka, Tokyo 181-8588, Japan}
\affiliation{Earth-Life Science Institute, Tokyo Institute of Technology, Japan}

\author{Tomoyasu Yamamuro}
\affiliation{Optcraft, Sagamihara, Kanagawa 252-0144, Japan}



\begin{abstract}
The direct measurement of the Universe's expansion history and the search for terrestrial planets in habitable zones around solar-type stars require extremely high-precision radial velocity measures over a decade. This study proposes an approach for enabling high-precision radial velocity measurements from space. The concept presents a combination of a high-dispersion densified pupil spectrograph and a novel telescope line-of-sight monitor. The precision of the radial velocity measurements is determined by combining the spectrophotometric accuracy and the quality of the absorption lines in the recorded spectrum. Therefore, a highly dispersive densified pupil spectrograph proposed to perform stable spectroscopy can be utilized for high-precision radial velocity measures. A concept involving the telescope line-of-sight monitor is developed to minimize the change of the telescope line-of-sight over a decade. This monitor allows the precise measurement of a long-term telescope drift without any significant impact on the Airy disk when the densified pupil spectra are recorded. We analytically derive the uncertainty of the radial velocity measurements, which is caused by the residual offset of the line-of-sights at two epochs. We find that the error could be reduced down to approximately 1 $cm/s$, and the precision will be limited by another factor (e.g., wavelength calibration uncertainty). A combination of the high precision spectrophotometry and the high spectral resolving power could open a new path toward the characterization of nearby non-transiting habitable planet candidates orbiting late-type stars. We present two simple and compact high-dispersed densified pupil spectrograph designs for the cosmology and exoplanet sciences. 

\end{abstract}

\keywords{techniques: radial velocities --- techniques: spectroscopic --- cosmology: cosmological parameters --- planets and satellites: terrestrial planets}


\section{Introduction} \label{sec:intro}

\cite{Struve+1952} proposed the performance of high-dispersion spectroscopy to search for unseen companions (i.e., exoplanets) by detecting the periodic change in the radial velocity of a star due to the Doppler effect. After improving the precision of the radial velocity measurements for over 20 years, a Jupiter-mass companion with an orbital period of approximately 3.5 days around Pegasi 51 was finally successfully discovered with a high-dispersion ELODIE spectrograph at the Haute-Provence Observatory in 1995 \citep{Mayor+1995}. About a quarter of 4000 exoplanets confirmed so far have been detected by the radial velocity measurements based on an exoplanet database\footnote{http://exoplanet.eu/}. This has rapidly expanded the research field's focus to search for life activity on exoplanets by characterizing the terrestrial planet atmospheres in the habitable zones around various types of stars. Several candidates found around nearby late-type stars, such as Proxima Centauri, Teegarden, and Trappist-1, have been confirmed by the radial velocity and transit photometry measurements \citep[e.g.,][]{Anglada-Escude+2016, Zechmeister+2019, Gillon+2016}, and transit spectroscopy and high-contrast imaging will characterize the atmospheres in the next two decades. Thanks to the low contrast ratio between a planet and its host star in the mid-infrared regime, the multi-band photometry and spectroscopy characterize the atmospheres of non-transiting terrestrial planets in the habitable zones around nearby late-type stars \citep{Kreidberg+2016,Snellen+2017,Fujii+2021}. In contrast, no candidates around Sun-like stars have been detected because of stricter requirements. The detection of candidates requires maintaining the precision of an order of 10 $cm/s$ over a few years. The precision of the current radial velocity planet-search instruments, including HARPS \citep{Pepe+2002} and HIRES \citep{Vogt+1994}, is limited to approximately 1 $m/s$ over a few years before 2016 \citep{Fischer+2016}. However, the next-generation radial velocity programs, such as the Echelle Spectrograph for Rocky Exoplanets and Stable Spectroscopic Observations\citep[ESPRESSO:][]{Pepe+2014, GonzalezHernandez+2018} and the Extreme Precision Spectrograph \citep[EXPRES:][]{Jurgenson+2016}, could achieve the precision of a few tens of $cm/s$ on the sky over a timescale of a few hours to a few days \citep[e.g.,][]{Petersburg+2020, Pepe+2021}. Thanks to advanced modal noise reduction techniques, advanced radial velocity instruments mostly overcame the continuous change in the atmospheric seeing conditions \citep{Mahadevan+2014, Halverson+2015, Sirk+2018, Petersburg+2018}. Note that the modal noise in a multi-mode fiber indicates coherent speckles at the output end of the fiber induced by optical interference \citep[e.g.,][]{Lemke+2011, Oliva+2019}. In addition, a precise wavelength calibration using the laser frequency comb \citep{Wilken+2012, Molaro+2013} compensates for a shift of the relative position between the spectrograph and the detector. In addition to the ground-based planet-survey programs, a simpler approach for space-based observatories has been investigated as a probe mission concept study of the Astro2020 Decadal Survey \citep{Plavchan+2019}. This approach can avoid a fundamental limitation caused by the telluric atmosphere, corresponding to $\sim$ 10 - 30 $cm/s$ in visible, and mitigate the stellar jitter with a wide observing bandwidth ranging from UV to near-infrared. A compact high-dispersion spectrograph suitable for space observatories could be realized by combining a Virtually Imaged Phased Array (VIPA) originally proposed for telecommunication \citep{Shirasaki+1996} as an Echelle grating with a general grating as a cross-disperser \citep{Bourdarot+2018, Zhu+2020}. The large dispersion angle of the VIPA can keep the size of the spectrograph with a resolving power of more than 100,000 below 1 $m$ long. However, the radial velocity measurements on both the ground- and space-based observatories will be eventually limited by the stellar jitter caused by an inhomogeneous surface with spots and plagues, which makes difficult to detect habitable planet candidates around Sun-like stars. Based on the long-term monitor of the Sun over three years \citep{Dumusque+2015, Dumusque+2021}, the Sun's radial velocity changes at the level of 1 $m/s$ from day to day.

\cite{Loeb+1998} proposed the application of existing high-precision radial-velocity instruments developed for exoplanet search to constraints of the cosmological parameters, such as the density parameters of the universe and the Hubble constant, developing an idea on the direct measurement of its global dynamics \citep{Sandage+1962}. Two spectroscopic observations with a time separation of 10 years would possibly allow us to detect the small redshift drifts of extragalactic objects if the signal-to-noise ratio of the cosmic signal is statistically proportional to the square root of the number of the observed astronomical objects. Neutral hydrogen atoms (HI) between the object and the Earth make a number of the Lyman alpha absorption lines in the shorter wavelength range than the Lyman alpha break (i.e., Lyman alpha forest); hence, quasars (hereafter referred to as QSOs) as one of the brightest extragalactic objects are a suitable background source for this purpose. Many absorption lines in the Lyman alpha forest improve the radial velocity precision. This idea was first applied to the HI 21 $cm$ absorption line in the continuum radio spectrum of a QSO, namely 3C286 \citep{Davis+1978}. The redshift drift of the HI 21 $cm$ again became the focus, thanks to the emergence of the Super Kilometer Array (SKA) \citep{Darling+2012}. The advantage of these redshift drift measurements is looking over the dynamical evolution of the universe without the cold dark matter model as the standard cosmological scenario. In contrast, the other measurements, including the cosmic microwave background \citep[CMB:][]{Sachs+1967, Smoot+1992}, the baryon acoustic oscillation \citep[BAO:][]{Sunyaev+1970,Peebles+1970}, and the Type Ia supernova surveys \citep{Goobar+1995}, require the standard cosmological model. Although the standard cosmological model has been widely accepted by successfully describing the universe structure at different cosmic epochs, a significant ($> 4 \sigma$) discrepancy between the Hubble constants derived from the measurements of the early and late universe, called the Hubble tension, was recently revealed \citep[e.g.,][]{Riess+2016,Verde+2019}. The standard cosmological model may require modification and new physics \citep[e.g.,][]{DiValentino+2021}. The initial condition of the CMB could not be extrapolated to its entire history; thus, even though the other methods successfully constrained the cosmological parameters and the Hubble constant \citep[e.g.,][]{Perlmutter+1999,Spergel+2003,Spergel+2007,Eisenstein+2005}, it is valuable to directly measure them from the redshift drift of an object without the $\Lambda$CDM model. Focusing on the fact that the Lyman alpha forest is produced by nutral hydrogen medium between the QSO and the observer, we can trace the expansion history of the Universe (i.e., redshift drifts at various redshifts) from an observation of a QSO, which may lead to solve the Hubble tension and discover new physics. 

For this purpose, the COsmic Dynamic and EXo-earth experiment (CODEX) to be mounted on the European Extremely Large Telescope (E-ELT) aims to reach an order of a few $cm/s$ over a decade in the era of extremely large telescopes \citep{Pasquini+2008}. Note that the project name of CODEX was changed to HIRES after the two science communities for the optical and near-infrared high-resolution spectrographs were merged. The HIRES instrument could restrict the cosmological parameters if the photon noise mainly limits the measurement accuracy of the redshift drift \citep{Liske+2008, Maiolino+2013}. However, the redshift drifts of QSOs at $z = 2 - 5$ over a decade are expected to be only a few $cm/s$ based on the latest results from the Planck satellite and the Sloan Digital Sky Survey \citep{Planck+2020,eBOSS+2020}. In other words, the systematic error of the radial-velocity measurements should be reduced down to the same level, as discussed in Section \ref{sec:basics}.

The uncertainty of the radial velocity measurements was first studied under the assumption that the spectrograph can achieve a photon-noise-limited performance \citep{Connes+1985,  Bouchy+2001, Figueira+2018}. In parallel, radial velocity planet-search programs revealed that not only the systematic noises, but also the random noises, such as the photon and detector noises, limit the precision of the radial velocity measurements \citep{Tuomi+2013}. While the measurement errors of the radial velocity instruments are reduced in proportion to the signal-to-noise ratio in its low regime, these errors are almost constant in its high regime \citep{Fischer+2016}. Following these studies, \cite{Plavchan+2015} revealed what types of systematic noises on fiber-fed high-dispersion spectrographs placed in a temperature-controlled room \citep[e.g.,][]{Mayor+2003} affect the radial velocity measurements. Each instrumental noise has been quantitatively evaluated \citep[e.g.,][]{Halverson+2016, Blackman+2020}. Next-generation radial velocity instruments could reduce the noise down to an order of 10 $cm/s$, thanks to the relevant technological advancements \citep[e.g.,][]{Wilken+2012,Mahadevan+2014,Halverson+2015}. While terrestrial planets in the habitable zones of nearby G-type stars could be found in the next decade, whether the Doppler drifts of the QSOs at $z = 2 - 5$, corresponding to a few $cm/s$ over a decade, are measured is still an open question. Even if the next-generation instruments can achieve a photon-noise-limited performance, different observatories should independently validate the measured cosmological parameters. HIRES is the only instrument designed for this purpose.

Here, we notice that the radial velocity measurements have an important similarity to transit spectroscopy. Two types of methods obtain the difference between the spectra recorded at two epochs. Considering that transit spectroscopy is directly limited by the photometric precision of the spectrograph, there is also a relation between the precision of the radial velocity measurements and the photometric precision. A densified pupil spectrograph concept proposed for transit spectroscopy of exoplanets from space \citep{Matsuo+2016} could be applied to the radial velocity measurements. The densified pupil spectrograph achieves the high precision spectrophotometry by producing the spectra of multiple sub-pupils on the detector plane, to which the primary mirror is optically conjugated. Low-order aberrations, such as the telescope line-of-sight jitter (i.e., pointing stability) and the primary mirror deformation, do not, in principle, change the spectral positions and sizes on the detector. On the other hand, the main difference between the radial velocity measurements and transit spectroscopy is the resolving power imposed on the spectrograph. While transit spectroscopy reduces the resolving power due to a decrease in the shot noise for each spectral element, the radial velocity measurements require a high resolving power for resolving a number of the absorption lines recorded in the spectrum. Therefore, we propose a high-dispersed densified pupil concept for enabling the high-precision radial velocity measurements. Furthermore, we notice that a combination of the high photometric precision and the high spectral resolving power could measure the Doppler shift of the planet spectrum embedded in the bright stellar light. As the systematic error is smaller, the stellar light gives a smaller impact on the planet spectrum. The high photometric precision allows us to observe planetary systems with high planet-to-star contrast ratios. It means that a new approach toward characterizing the reflected light of nearby non-transiting potential habitable planets can be developed without a high angular and high contrast technique. Thus, the high-dispersion densified pupil spectrograph concept may be useful not only for enabling the high-precision radial velocity measurements but also for characterizing the atmospheres of nearby habitable planet candidates.

However, we must consider the new impact of the telescope line-of-sight drift on the photometric stability in visible and near-infrared regimes. While the surface figure errors of the dispersive element placed on the focal plane could be regarded as negligible at the mid-infrared wavelengths, the speckles caused by the wavefront errors on the detector plane should be considered at shorter wavelengths. The line-of-sight jitter changes the image position on the dispersive element; thus, various speckle patterns are formed during observations. Finally, the drift of the telescope line-of-sight during one observation and its difference between two observations with a period of a decade will limit the accuracy of the radial velocity measurements. We introduce herein a new sensor concept for measuring the image position on the dispersive element. This sensor could precisely work without any significant impact on the point spread function (PSF) core at the same time when the densified pupil spectra are recorded. This sensor is different from general slit viewers placed on the spectrograph entrance \citep[e.g.,][]{Rayner+2003,Iseki+2008}. The photon-noise-limited precision of the radial velocity measurements may be possible by combining the densified pupil spectrograph with this type of sensor. 

This study proposes a new approach for achieving high-precision radial velocity measurements over a decade and characterizing the atmospheres of nearby habitable planet candidates on future space observatories. This concept is mainly composed of two sub-systems: 1) a high-dispersion densified pupil spectrograph; and 2) a telescope line-of-sight monitor. In Section \ref{sec:basics}, we derive the precision of the radial velocity measurements and summarize the requirements for enabling the photon-noise-limited performance. Section \ref{sec:concept} presents an overview and the limitation based on an analysis of the wavefront propagation through the spectrograph. Sections \ref{sec:evaluation} and \ref{sec:discussion} show the optical designs of the high-dispersion spectrographs optimized for direct measurement of the universe's expansion history and characterization of the atmospheres of nearby habitable planet candidates orbiting late-type stars, respectively. While the former spectrograph has the resolving power of approximately 10,000 in the optical wavelength regime, the resolving power of the latter one could reach to 70,000 over the wavelength range of 750 to 980 $nm$. We investigate the possibilities of whether the two science cases are realized with this approach. Finally, we discuss why this proposed approach could be beneficial for space telescopes, comparing this concept with ground-based fiber-fed spectrographs for the Doppler planet search in Section \ref{sec:hds_for_space_telescopes}.

\section{Basics of radial-velocity measurements} \label{sec:basics}
Before presenting the concept overview, we will explain why the systematic noise floor should be considered for enabling high-precision radial velocity measurements. In this section, we derive the uncertainty of the radial velocity measurements caused by the shot noise and the systematic instrumental noise. We also discuss the requirements for enabling the photon-noise-limited performance and compare them with those of transit spectroscopy. 

\subsection{Ideal limitation} \label{subsec:ideal_limitation}
According to the previous studies \citep{Connes+1985,Bouchy+2001,Figueira+2018}, the radial velocity signal for the $i$-th spectral element is
\begin{equation}
	\label{eqn:signal_ith_element}
	\frac{\delta V}{c}(i) = \frac{N_{2}(i)-N_{1}(i)}{\lambda(i) \left(\frac{\partial N_{true}}{\partial \lambda} (i) \right)},
\end{equation}
where $\lambda$ is the observation wavelength; $N_{1}(i)$ and $N_{2}(i)$ are the numbers of the observed photons for the $i$-th spectral element at the two epochs 1 and 2, respectively; and $N_{true}(i)$ is the true spectrum of a target for the $i$-th spectral element. Each spectral element has a different error, the estimation of the radial velocity is written as an weighted average of the radial-velocity signals over the entire wavelength range:
\begin{equation}
	\label{eqn:estimation}
	\left(\frac{\delta V}{c}\right)_{wav} = \frac{\sum_{i} (\frac{\delta V}{c}(i) W(i))}{\sum_{i} W(i)},
\end{equation}
where $W (i)$ shows the optimum weight function for the $i$-th spectral element. The weighted average is equal to the maximum likelihood estimate under a condition that the optimum weight function is proportional to the inverse square of the standard deviation of each measurement value:
\begin{equation}
	\label{eqn:W_i}
	W(i) = \frac{1}{\sigma^{2}(i)},
\end{equation}
where $\sigma (i)$ is the random noise attached to the radial velocity signal of the $i$-th spectral element. Given that there is no systematic effect originated from the telescope and instrument, the signal variation is only caused by the shot noise, $\sigma_{shot}$; this assumption corresponds to the fundamental limitation case. Under this assumption, the variance of the weighted average, $\sigma^{2}_{wav}$, which represents the square of the uncertainty of the radial-velocity measurements, is presented as
\begin{eqnarray}
	\label{eqn:sigma_wav}
	\sigma^{2}_{wav} &=& \frac{1}{\sum_{i} W(i)} \nonumber \\
	&=& \frac{1}{\sum_{i} \frac{1}{\sigma^{2}_{shot}(i)}},
\end{eqnarray}
where we assumed that the spectrum slope is not changed during the two epochs. Given that the shot noise is independently generated at each epoch, the root-mean-square of the difference between the observed data at the two epochs, $[N_{2}(i) - N_{1}(i)]_{RMS}$, is $\sqrt{2N_{true}(i)}$; hence, $\sigma^{2}(i)$ is written as follows: 
\begin{equation}
	\label{eqn:sigma_i}
	\sigma^{2}(i)=\frac{2N_{true}(i)}{\lambda(i)\left(\frac{\partial N_{true}}{\partial \lambda}(i) \right)}.
\end{equation}
The weighted average variance under the fundamental limitation case is rewritten as
\begin{eqnarray}
	\label{eqn:ideal_limit}
	\sigma^{2}_{wav} &\simeq & \frac{2}{\sum_{i} \frac{\lambda^{2}(i)}{N_{true}(i)}\left(\frac{\partial N_{true}}{\partial \lambda} (i)\right)^{2}} \nonumber \\
 &=& \frac{2}{Q_{photon}^{2} \sum_{i} N_{true}(i)}, 
\end{eqnarray}
where $Q_{photon}$ is a dimensionless quantity that represents how much the obtained spectrum is suitable for the radial velocity measurements in terms of the spectral shape. The uncertainty of the radial velocity measurements is decreased by the square root of the number of the photo-electrons collected over the entire wavelength range, $\sqrt{\sum_{i}N_{true}(i)}$. The $Q_{photon}$ factor is defined as follows: 
\begin{equation}
	\label{eqn:Q_original}
	Q_{photon} \equiv \frac{\sqrt{\sum_{i} \frac{\lambda^{2}(i)}{N_{true}(i)}\left(\frac{\partial N_{true}}{\partial \lambda}(i)\right)^{2}}}{\sqrt{\sum_{i} N_{true}(i)}}. 
\end{equation}
The $Q_{photon}$ factor defined above is equal to the original $Q$ specified by the previous studies \citep{Connes+1996,Bouchy+2001}. Note that, if a change in the spectrum slope at the two epochs is not negligible, the change contributes to the weighted average variance and the $Q_{photon}$ factor \citep{Liske+2008}. The $Q_{photon}$ factor strongly depends on the resolving power and the intrinsic spectrum of the target. A higher resolving power resolves the spectral feature more; hence, the $Q_{photon}$ factor generally increases as the resolving power is higher. The uncertainty of the radial velocity measurements for the photon-noise-limited case is determined by the $Q_{photon}$ factor and the number of the photons collected over the observation bandwidth, $\sum_{i} N_{true}(i)$. The uncertainty is inversely proportional to the square root of the integration time, which is the same as the relationship between the signal-to-noise ratio and the integration time for general observations. 

\subsection{Impact of systematic noise on the radial velocity measurements} \label{subsec:systematic_limitation}

In reality, additional random signal variations caused by the detector noise and the random line-of-sight jitter exist. These signal variations increase the variance of each spectral element, $\sigma^{2}$; thus, the integration time required for reduction to the same variance level is longer than that for the fundamental limitation case. Furthermore, some systematic signal variations originate from unknown systematic effects, such as the change of the detector gain and the telescope line-of-sight drift (i.e., low-frequency pointing error). The static error cannot be smoothed out, even by averaging the obtained data over a long term; therefore, the measurement precision would be finally limited by the static systematic noises. Given that the systematic noise distribution is a Gaussian function centered at 0 along the spectral element, the variance of each spectral element, $\sigma^{2}$, can be written as the sum squared of the random and systematic noises:
\begin{equation}
	\label{eqn:tot_sigma}
	\sigma^{2}(i,t_{integ}) = \sigma^{2}_{shot}(i,t_{integ}) + \sigma^{2}_{detector}(t_{integ}) + \sigma^{2}_{jitter}(i,t_{integ}) + \sigma^{2}_{static}(i),
\end{equation}
where $\sigma_{detector}$ and $\sigma_{jitter}$ are the signal variations caused by the random detector noises and the telescope line-of-sight jitter, respectively, $\sigma_{static}$ is a static noise caused by the telescope pointing drift and the systematic change in the detector gain, and $t_{integ}$ is the integration time. All noises, except for the detector noises, have different wavelength dependencies. As will be discussed in Section \ref{sec:concept}, the signal variations induced by the line-of-sight jitter and drift originate from a partial loss of the amplitude spread function (ASF) on the entrance slit of the spectrograph (hearafter referred as to the "motion loss"). 

All noises, except for the static noises, could be roughly reduced by the squared root of the integration time, $t_{integ}$; hence, the standard deviation of each spectral element becomes the static systematic noise under the condition that the integration time is infinite. Given that the systematic noise floor of the $i$-th spectral element is $\epsilon (i)$, the systematic error of the difference between the observed data at two epochs limited by the static noise is approximately written as $\epsilon (i) N_{true}(i)$. When the integration time, $t_{integ}$, goes to infinity, the weighted average variance is reduced to
\begin{equation}
	\label{eqn:limitation_system}
	\lim_{t_{integ} \to \infty} \sigma^{2}_{wav} \simeq \frac{1}{\sum_{i} \frac{1}{\sigma^{2}_{static}(i)}}.
\end{equation}	
The static noise for the $i$-th spectral element is derived as follows through its error propagation:
\begin{equation}
	\sigma_{static}(i) = \frac{\partial \left(\frac{\delta V}{c} (i) \right)}{\partial u (i)} (u(i) - \overline{u}),
\end{equation}
where $u(i)$ shows the difference between the numbers of the photons collected at two epochs for the $i$-th spectral element, $N_{2}(i) - N_{1}(i)$, and $\overline{u}$ is its average. The radial velocity signal variance can be rewritten with Equation \ref{eqn:signal_ith_element} as follows because the systematic error of $u(i) -\overline{u}$ is defined as $\epsilon(i) N_{true}(i)$: 
\begin{equation}
	\label{eqn:limitation_system_2}
	\lim_{t_{integ} \to \infty} \sigma^{2}_{wav} = \frac{1}{\sum_{i} \left( \frac{1}{\epsilon(i)} \frac{\lambda(i)}{N_{true}(i)}\left(\frac{\partial N_{true}}{\partial \lambda}(i)\right) \right)^{2} }.
\end{equation}
The statistics of the static systematic noise is independent of that of the object spectrum. $\left(\frac{\lambda(i)}{N_{true}(i)}\frac{\partial N_{true}}{\partial \lambda}(i)\right)^{2}$ is roughly close to a log-normal distribution. When the static noise is Gaussian distribution along the wavelength, the right-hand side of Equation \ref{eqn:limitation_system_2} is more than a few times smaller than a simple multiplication of $\frac{1}{\sum_{i} \left(\frac{\lambda(i)}{N_{true}(i)}\left(\frac{\partial N_{true}}{\partial \lambda}(i)\right) \right)^{2}}$ and $\bar{\epsilon}$, where $\bar{\epsilon}$ is the average of the systematic noise over the entire wavelength range. This is because $\epsilon(i)$ close to 0 largely increases the denominator in the right-hand side of Equation \ref{eqn:limitation_system_2}. We numerically confirmed it, using the simulation spectra of the Lyaman alpha forest and the Sun-like star, which will be introduced later in this subsection. In contrast, if a strong correlation between the static noises of the different spectral elements exists, the uncertainty of the measurements could be larger than for the no-correlation case. The static noise tends to originate from the telescope line-of-sight drift and the change in the detector gain; therefore, the systematic noise errors are correlated to some extent. When the static noise has the same value of $\bar{\epsilon}$ over the whole wavelength range, the radial velocity signal variance is maximized:
\begin{equation}
	\label{eqn:limitation_system_latter_case}
	\lim_{t_{integ} \to \infty} \sigma^{2}_{wav} \leq \frac{\bar{\epsilon}^{2}}{Q_{sys}^{2}},
\end{equation}	
where $Q_{sys}$ is
\begin{equation}
	\label{eqn:Q'_factor}
	Q_{sys} = \sqrt{\sum_{i} \left(\frac{\lambda(i)}{N_{true}(i)}\frac{\partial N_{true}}{\partial \lambda}(i)\right)^{2}}.
\end{equation}
The systematic noises for all of the spectral elements contribute to the variance of $\sigma_{wav}^{2}$ in the above equation; hence, the above assumption corresponds to the worst scenario for the radial velocity measurements.

\begin{figure}
	 \centering
	\includegraphics[scale=0.1,height=7cm,clip]{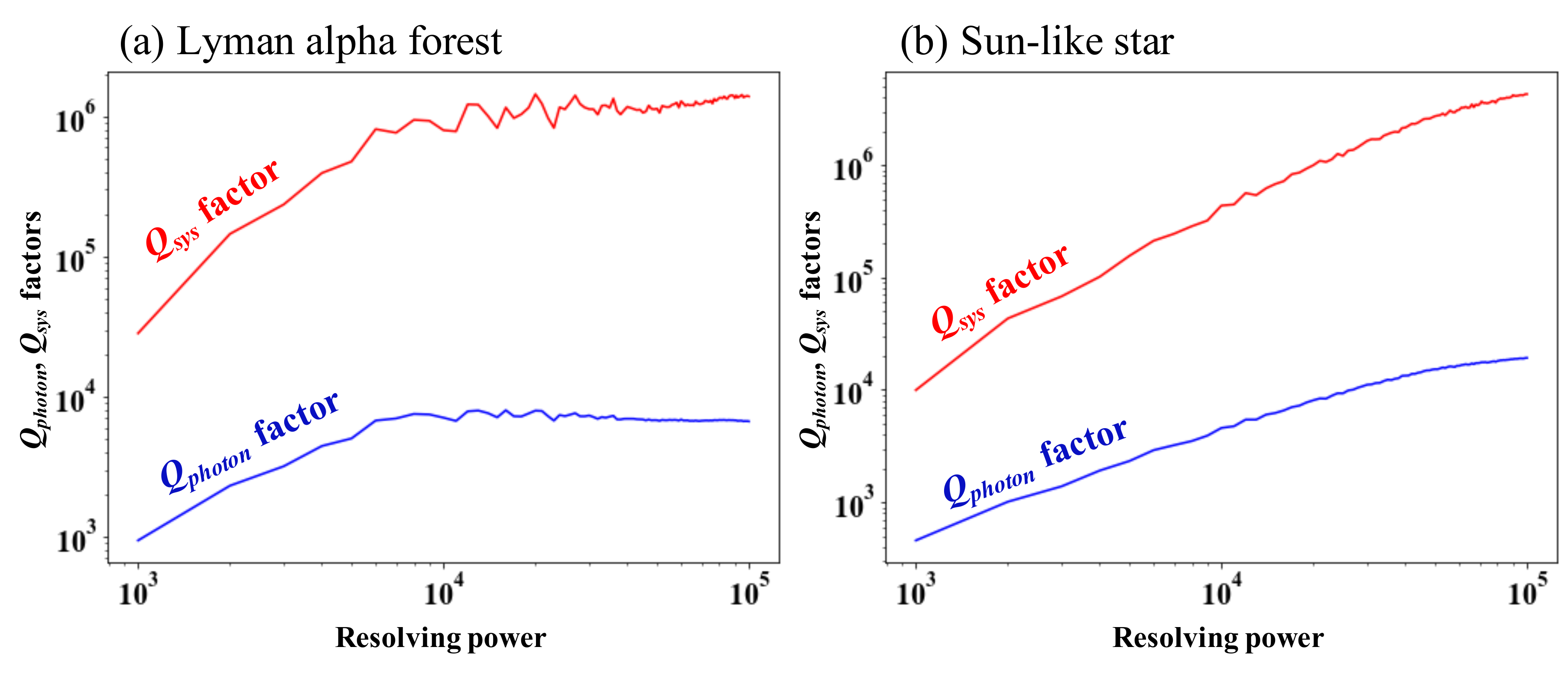}
	\caption{$Q_{photon}$ and $Q_{sys}$ factors as a function of the resolving power for (a) the Lyman alpha forest of a QSO at $z = 3$ and (b) the spectrum of the Sun-like star in optical, respectively. The blue and red lines show the $Q_{photon}$ and $Q_{sys}$ factors, respectively. The wavelength ranges of the two cases are approximately (a) 450 - 480 $nm$ and (b) 450 - 550 $nm$, respectively. Section \ref{sec:appendix} describes in detail the simulation of the Lyman alpha forest spectra. }
	\label{fig:Q_factor}
\end{figure}

$Q_{sys}$ is also a dimensionless quantity that almost looks the same as the original $Q_{photon}$ shown in Equation \ref{eqn:Q_original}. However, the two values are greatly different. Figure \ref{fig:Q_factor} shows the $Q_{photon}$ and $Q_{sys}$ factors as a function of the resolving power for two spectra: a Lyman-alpha forest of a QSO at $z=3$ in optical and a spectrum of a Sun-like star in optical. The Lyman alpha forest spectra used in this study were generated from the Illustris-TNG simulation\footnote{\url{https://www.tng-project.org/}} \citep{Springel:2018, Marinacci:2018, Naiman:2018, Nelson:2018, Pillepich:2018b} using the \texttt{$fake\_spectra$} \footnote{\url{https://github.com/sbird/fake_spectra}} package \citep{Bird:2014a}. Appendix \ref{sec:appendix} describes the details of the simulation parameters. The spectrum of a Sun-like star with the surface gravity of 4.5 $log(cm/s^{2})$, effective temperature of 5800 $K$, and solar metallicity was generated by the \texttt{$BT-settle$} model \citep{Allard+2012} in the theoretical spectra web server\footnote{http://svo2.cab.inta-csic.es/theory/newov2/}. The difference of the two values became larger as the resolving power increased. The number of the spectral elements was almost proportional to the resolving power; thus, the ratio of $Q_{sys}$ to $Q_{photon}$ almost increased with the square root of the resolving power (Figure \ref{fig:Q_factor}). 

\begin{figure}
	 \centering
	\includegraphics[scale=0.1,height=7cm,clip]{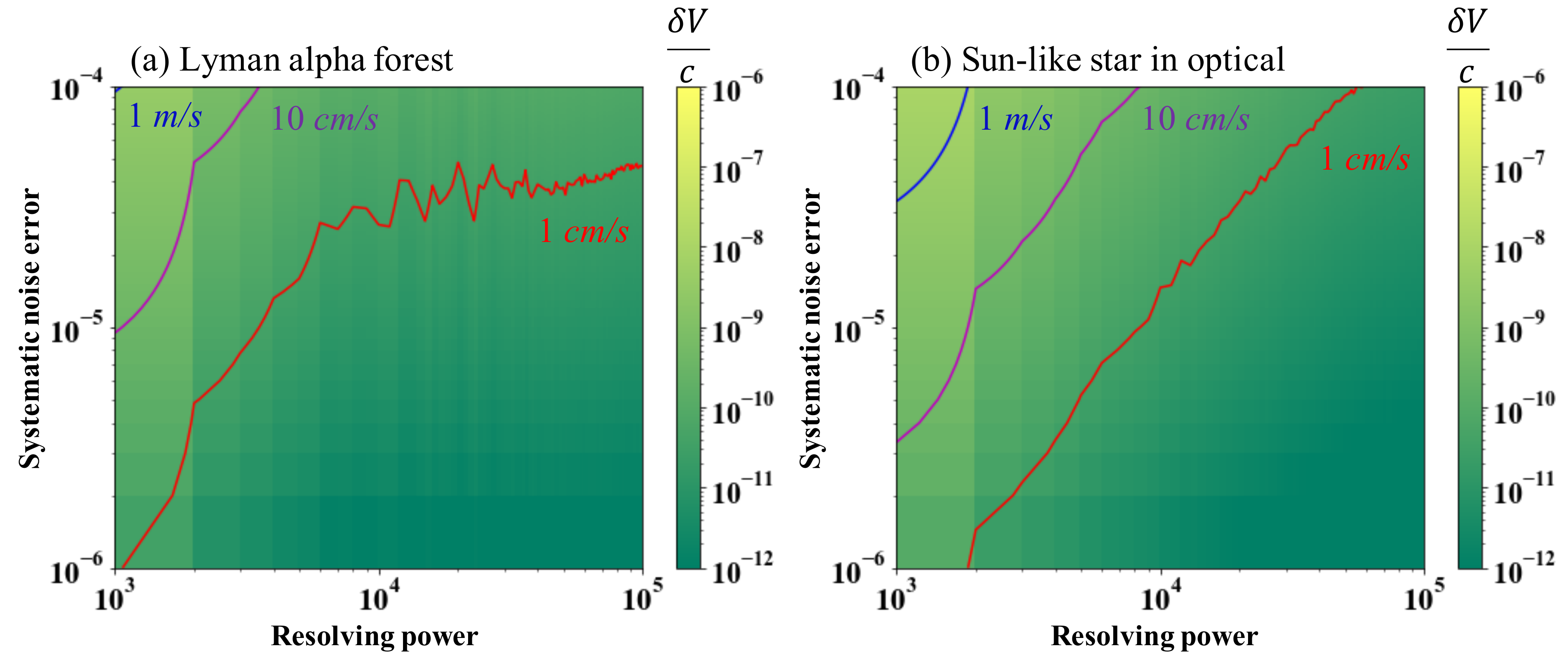}
	\caption{Uncertainty of the radial-velocity measurements caused by the systematic noise error for (a) the Lyman alpha forest of a QSO at $z = 3$ and (b) the spectrum of the Sun-like star in optical as a function of the resolving power, respectively. The red, purple, and blue lines represent the precisions of $1 cm/s$, $10 cm/s$, and $1 m/s$, respectively. }
	\label{fig:uncertainty_noise_floor}
\end{figure}

Figure \ref{fig:uncertainty_noise_floor} shows the uncertainty of the radial velocity measures caused by the systematic errors for the worst case. The measurement uncertainty was determined by a combination of the resolving power and the systematic noise error. The $Q_{sys}$ factor became larger as the resolving power became higher (Figure \ref{fig:Q_factor}); therefore, the instrumental noise floor requirement was mitigated more for the higher resolving power. To achive 1 $cm/s$ precision, the noise floor should be reduced down to a few and a few tens of parts-per-million ($ppm$) for the resolving powers of 10,000 and 100,000, respectively. Although the precision of actual radial velocity measurements would lie above the worst scenario in reality, the systematic noise floor requirement should be equal to that derived based on the worst sceanrio for reliable measurements because the distribution of the systematic errors along the wavelength is unknown. In this paper, we applied Equation \ref{eqn:limitation_system_latter_case} as the radial velocity measurement unceratinty caused by the systematic noise errors. 

A change in the intrinsic spectrum of a target at two epochs could become another systematic effect. In this study, we neglect the change in the intrinsic spectrum over a decade. While spots and faculae on the surface limit the precision of the radial velocity measurements in optical \citep[e.g.,][]{Makarov+2009}, the systematic effects would be largely reduced at longer wavelengths \citep{Marchwinski+2015}. The assumption is also appropriate for the direct measurement of the cosmological parameters because a number of objects are included in each QSO \citep{Loeb+1998}. 

\subsection{Similarities and differences between radial velocity and transit spectroscopy}\label{subsec:rv_transit}
The signal of the radial velocity measures is a change in spectra recorded at two epochs along the wavelength direction; hence, high-precision radial velocity measurements require not an absolute, but a relative photometric precision. A spectrograph will reach the photon-noise-limited performance in terms of the radial velocity measures under the assumption that the observation and the instrument conditions at the two epochs are perfectly the same. Conversely, when the conditions are different, the precision of the radial velocity measurements would be limited by unknown systematic effects. The transit spectroscopy signal is the difference between the spectra at in- and out-of-transits; thus, both the radial velocity measurements and transit spectroscopy need a similar instrument concept to improve their performances. The key parameter of the instrument concept is "stability," which is defined herein as the ability of keeping an instrument condition at two epochs. A high stability indicates that the conversion factor of the incident photons (to the primary mirror) to the digitalized data is constant at two epochs. 

Based on the considerations in the previous subsection, the following two conditions are required to achieve the photon-noise-limited performance. 1. The random signal variation should be mainly determined by the photon noise: 
\begin{equation}
	\sigma_{shot}(i,t_{integ}) > \sqrt{\sigma^{2}_{detector}(t_{integ}) + \sigma^{2}_{jitter}(i,t_{integ})}.
\end{equation}
2. The uncertainty of the radial velocity measurements induced by the systematic effects should be smaller than the signal variation due to the shot noise:
\begin{equation}
\frac{\bar{\epsilon}}{Q_{sys}} < \frac{1}{Q_{photon}\sqrt{\sum_{i}N_{true}(i)}}.
\end{equation}
These conditions are similar to those imposed for transit spectroscopy for characterizing the exoplanet atmospheres. 

Another key parameter for the two observing methods is the resolving power of the spectrograph. The radial velocity and transit spectroscopy require different resolving power. While transit spectroscopy demands relatively low resolving power to reduce the shot noise for each spectral element, the radial velocity measurement requires a higher resolving power to enhance the $Q_{photon}$ and the newly defined $Q_{sys}$ factors. The higher $Q_{sys}$ could mitigate the requirement on the systematic noise floor. In addition, the time-spans for maintaining the photon-noise-limited performance (i.e., photometric stability) are different for the two cases. Transit spectroscopy, including the phase curve measurements, typically spans from a few hours to a few days, while the radial velocity measurements for the search for Earth-twins orbiting Sun-like stars and the direct measurement of the Universe's expansion history require the time-spans of a few years to a decade. Thus, high-precision radial velocity measurements must simultaneously achieve both high-dispersion spectroscopy and photon-noise-limited performance over a time span of a few years to a decade. The requirements imposed on the radial velocity measurements are tougher than those for transit spectroscopy.

\section{Instrument concept} \label{sec:concept}
In this section, we propose an instrument concept for enabling high-precision radial velocity measurements based on the requirements discussed in Section \ref{sec:basics}. This concept is a combination of a high-dispersion densified pupil spectrograph and a novel telescope line-of-sight monitor.

\subsection{Overview}
\begin{figure}
	 \centering
	\includegraphics[scale=0.1,height=8cm,clip]{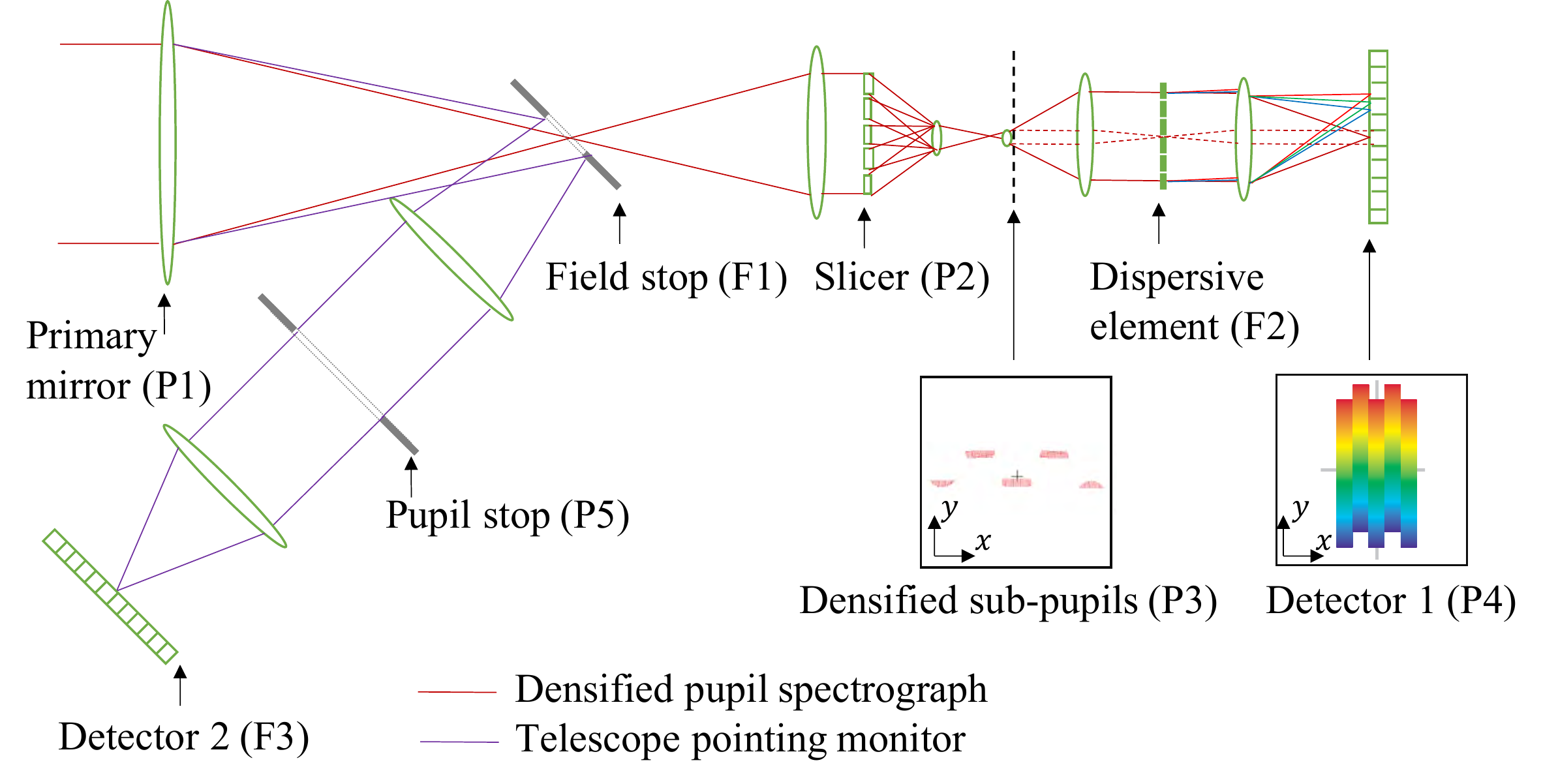}
	\caption{Conceptual diagram of a densified pupil spectrograph combined with a telescope line-of-sight monitor. The red and purple lines show the beams introduced to the densified pupil spectrograph and the line-of-sight monitor, respectively. The terms of "P" and "F" in parentheses represent the pupil and focal planes, respectively. The red dashed line in the spectrograph shows the beam propagation without a wave diffraction. The coordination system of the pupil plane is $(x,y)$. }
	\label{fig:concept_radial_velocimetry}
\end{figure}

The direct measurement of the Universe's expansion history  and the search for habitable planet candidates around Sun-like stars require a spectrograph that maintains the photon-noise-limited performance over a decade. We present a new possibility for realizing these science cases by developing an instrument concept that is complementary to widely common high-dispersion spectrographs. This instrument concept is mainly composed of two sub-systems: 1) a high-dispersion densified pupil spectrograph; and 2) a telescope line-of-sight monitor suitable for its spectrograph. Figure \ref{fig:concept_radial_velocimetry} shows an overview of this instrument concept. The light coming from the primary telescope is divided into two beams by a field stop, whose surface is coated by a reflective material. While the light inside the field stop is introduced to the densified pupil spectrograph, that outside the field stop is reflected by the surface and is introduced to the telescope line-of-sight monitor. The densified pupil spectrograph can minimize the impacts of the line-of-sight jitter and drift on the photometric precision, while the line-of-sight monitor reduces the change in the telescope line-of-sight over a long period. Although this line-of-sight monitor looks like a conventional slit viewer \citep[e.g.,][]{Rayner+2003,Iseki+2008} and a coronagraphic low-order wavefront sensor \citep[CLOWFS;][]{Guyon+2009} in terms of dividing the incident beam into two on the focal plane, the wavefront propagations through the two telescope pointing sensors are totally different, as discussed in Section \ref{subsec:sensor_drift}. In addition, the position of each spectral element is stable on the detector; thus, the spectrum shift caused by the Doppler effect may be able to be measured along the wavelength without any wavelength calibrators, such as a gas absorption cell and a Laser Frequency Comb (LFC), once the relation between the detector position and the wavelength is precisely determined. Thus, a combination of the densified pupil spectrograph and telescope line-of-sight monitor allows us to perform highly stable spectroscopy against the telescope pointing drift and its change over a decade. 

\begin{figure}
	 \centering
	\includegraphics[scale=0.1,height=6.3cm,clip]{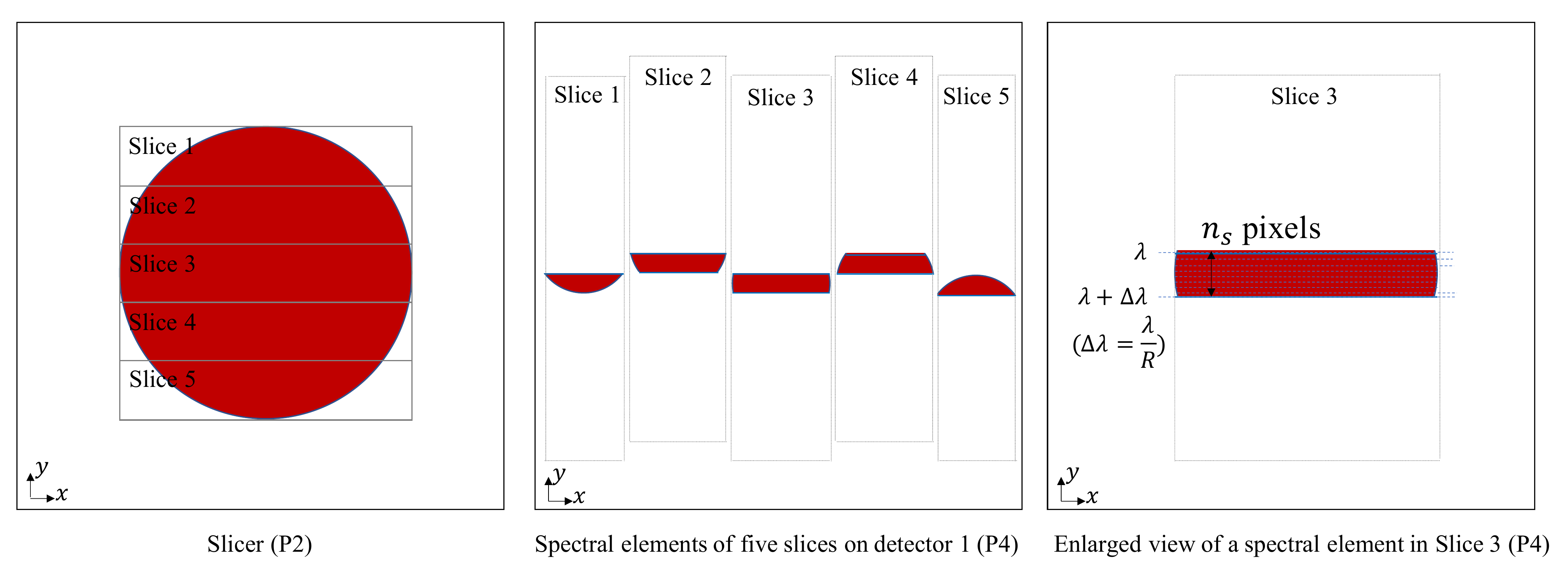}
	\caption{Conceptual diagram of the densified pupil spectra on the detector 1. The left panel shows a pupil image on the pupil slicer with five rectangular mirrors (P2). The red circle represents the entrance pupil on the P2 plane. The pupil is divided into five with the pupil slicer. The middle panel shows five densified pupil spectra on the detector 2 (P4). The five dotted rectangulars, named "Slice 1, 2, ..., 5," represent the five spectra formed on the P4 plane. The red arcs are spectral elements at a specific wavelength. The right panel shows an enlarged view of a spectral element in Slice 3. The difference of the wavelength from the edge to edge of the spectral element corresponds to $\Delta \lambda$. Each spectral element is sampled by $n_{s}$ detector pixels along the wavelength direction (i.e., $y$ direction). $n_{s}$ is determined such that the detector noise is smaller than the photon noise from an astronomical object. $n_{s}$ is more reduced as the resolving power is higher.}
	\label{fig:spectral_element}
\end{figure}

We will briefly introduce how the densified pupil spectrograph works and explain why the telescope line-of-sight monitor could improve the precision of the radial velocity measurements. The light introduced to the densified pupil spectrograph is first divided into several with a slice mirror composed of a few rectangular mirrors at the entrance pupil, to which the primary mirror is optically conjugated. Each divided beam is densified with two concave mirrors. Densified sub-pupils are then formed at the spectrograph entrance, corresponding to the pupil plane. The size of each densified sub-pupil is close to the wavelength; hence, each beam is diffracted as a point source from the spectrograph entrance. Each diffracted beam is collimated by a collimate mirror, and the ASF is formed on a dispersive element of the focal plane. Finally, the spectra of the densified sub-pupils are formed on the detector 1. Figure \ref{fig:spectral_element} shows the conceptual diagram of the densified sub-pupil spectra. We define the densified sub-pupil as one spectral element. Each spectral element is sampled by a higher number of detector pixels compared to that of a general spectrograph because of the division of the entrance pupil into several. Thanks to this feature, each spectral element is less affected by the bad pixels and the cosmic rays. The impact of the pixel-to-pixel quantum variation on the wavelength calibration with a calibration source can be largely mitigated, thanks to the larger detector samplings \citep{Goda+2018, Matsuo+2020}. The long-term stability of the calibration source will be a fundamental limitation on the wavelength calibration (see Section \ref{sec:evaluation}).

\begin{figure}
	 \centering
	\includegraphics[scale=0.1,height=7cm,clip]{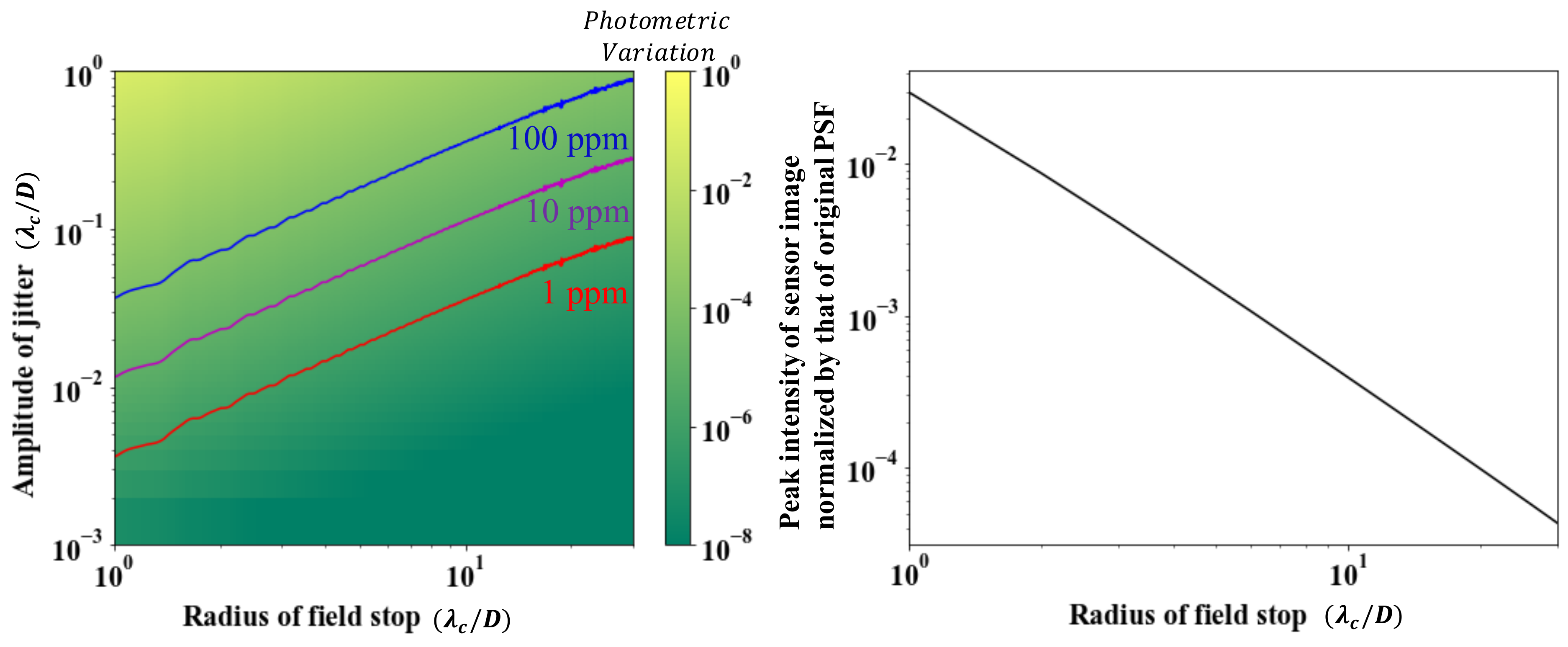}
	\caption{(Left) Averaged photometric variation caused by the motion loss over the observing bandwidth equivalent to the central wavelength, $\lambda_{c}$, as the functions of the amplitude of the telescope line-of-sight jitter and the field-stop radius, which are normalized by $\frac{\lambda_{c}}{D}$. The red, purple, and blue solid lines represent the photometric variation of 1, 10, and 100 $(ppm)$, respectively. (Right) The peak intensity of the image formed on the telescope line-of-sight monitor, which is normalized by that of the original PSF, as a function of the field-stop radius in the unit of $\frac{\lambda_{c}}{D}$.}
	\label{fig:motion_loss}
\end{figure}

The detector plane is optically conjugated to the primary mirror; therefore, any wavefront errors on the pupil plane, in principle, do not affect the spectra of the densified sub-pupils. However, the telescope line-of-sight jitter and drift introduce intensity variations even under the assumption that the optical system is ideal because a motion loss (i.e., a partial loss of the ASF) is generated on the field stop and the dispersive element, which are placed on the focal plane. According to the previous study of \cite{Itoh+2017}, the photometric variation caused by the motion loss on the focal plane for a circular isotropic aperture is written as
\begin{equation}
	\label{eqn:sigma_jitter}
	\sigma_{jitter} = \frac{\pi a}{2} \left| \frac{\partial}{\partial r} PSF(r)|_{r=a} \right| (\Delta \theta)^{2},
\end{equation}
where $r$ is the radial distance from the center of the field (i.e., $r = \sqrt{\xi^{2} + \eta^{2}}$); $PSF$ is the point spread function formed on the focal plane; $a$ is the field-stop radius in the radian unit; and $\Delta \theta$ represents a change of the telescope line-of-sight at two epochs normalized by the spatial resolution of a telescope. The photometric variation caused by the motion loss was mainly determined by a combination of the telescope line-of-sight jitter and the field-stop radius. The left panel of Figure \ref{fig:motion_loss} shows the averaged photometric variation caused by the motion loss in a bandwidth equivalent to the central wavelength. The line-of-sight jitter requirement can be more mitigated as the field-stop radius is larger. To achieve the photometric precision of 10 $ppm$, the jitter amplitude should be reduced to 0.01 and 0.1 times the spatial resolution for the radii of 1 and 10 times its resolution, respectively. In Section \ref{sec:evaluation}, we will derive the optimum field-stop radius, such that the photometric variation is smaller than 10 $ppm$.

However, even if the signal variation caused by the motion loss can be negligible, the telescope line-of-sight jitter and drift degrade the photometric precision via another intensity variation. The densified pupil spectrograph stabilizes the spectra of the densified sub-pupils on the detector plane because its detector is optically conjugated to the primary mirror. A dispersive element is inevitably placed on the focal plane, and the tilt in the phase of each densified sub-pupil on the P3 plane moves its Airy disk formed on the dispersive element. The wavefront aberration introduced on the dispersion element of the focal plane forms speckles on the detector plane. The speckles are brighter because the wavelength is shorter (i.e., in the visible and near-infrared regimes). As a result, a combination of the newly added wavefront aberration with the change in the telescope line-of-sight produces various speckle patterns on the detector during the observation. Focusing on the fact that the line-of-sight jitter randomly changes, the speckle patterns induced by the jitter are smoothed out for a long-term observation. Conversely, the intensity variation caused by an offset of the telescope line-of-sight at two epochs remains as a static noise corresponding to the systematic noise floor of the photometry, $\epsilon(i)$. Thus, the photometric variations induced by the telescope pointing drift could inevitably degrade the radial velocity measurements at the visible and near-infrared wavelengths. 

However, it is hard to measure the telescope line-of-sight drift on the densified pupil spectrograph because the detector plane is optically conjugated to the primary mirror. The telescope line-of-sight drift could not be precisely measured on general imaging spectrographs because the light coming from an astronomical object is dispersed and is blurred along the wavelength direction on the detector. In addition, a general slit viewer cannot measure the telescope pointing drift at the same time when the spectrum of an astronomical object is recorded because the ASF core is introduced to the densified pupil spectrograph. Based on this background, we employed a sensor for measuring the telescope line-of-sight, using the light blocked by the field stop (i.e., the ASF base). Thanks to the simultaneous measurements of the telescope pointing drift and the densified pupil spectra, we could reduce the impact of the long-term drift on the photometric precision by compensating it during the observation. This sensor would be suitable for measuring the slow change in the telescope line-of-sight because the amount of the photons introduced to the sensor is limited (right panel of Figure \ref{fig:motion_loss}). In contrast, if a bright object is observed with a large telescope, this monitor possibly works as a real-time sensor, such as the wavefront sensors used for adaptive optics \citep[e.g.,][]{Hardy+1998, Guyon+2005} and the low-order wavefront sensors for a coronagraph \citep{Guyon+2009}.

In the following subsections, we analytically investigate how the wavefront aberrations on the dispersive element affect the photometric precision based on an analysis of the wavefront propagation through the densified pupil spectrograph. We also introduce the telescope line-of-sight monitor, describing how the monitor precisely measures the telescope line-of-sight.

\subsection{Impact of long-term drift on radial velocity measurements} \label{subsec:impact_rv_measurements}

In this subsection, we analytically derive how the difference of the line-of-sights at two epochs affects the radial velocity measurements. No optical element exists on the focal plane, except for the field stop before the spectrograph entrance (i.e., the P3 plane); thus, the wavefront propagation from the entrance of the spectrograph (P3) to the detector 1 (P4) is focused.

The coordinate systems of the pupil and the focal planes were set to $(x,y)$ and $(\xi,\eta)$, respectively. $\eta$ and $y$ were parallel to the dispersion direction (i.e., wavelength one). When the wavefront aberration of each densified sub-pupil on the P3 plane, including the telescope pointing error, is $\phi_{P3} (x, y, t_{k})$ at the $k$-th epoch, the complex amplitude of its densified sub-pupil at the $i$-th epoch is written as $E_{P3}(x,y,\lambda(i),t_{k})=E_{0}(x,y,\lambda(i))\exp\left(i\frac{2\pi}{\lambda(i)}\phi_{P3}(x,y,t_{k})\right)$, where $E_{0}(x,y,\lambda(i))$ is the electric field of the entrance pupil from an observed astronomical object at the wavelength of $\lambda(i)$. The electric field of each densified sub-pupil on the $i$-th dispersive element at the $k$-th epoch is 
\begin{equation}
	\label{eqn:E_F2}
	E_{F2}(\xi,\eta,\lambda(i),t_{k}) = FT\left\{E_{0}(x,y,\lambda(i))\exp\left(i\frac{2\pi}{\lambda(i)}\phi_{P3}(x,y,t_{k})\right) \right\} E_{g}(\xi,\eta)\exp\left(i\frac{2\pi}{\lambda(i)}\phi_{g}(\xi,\eta,t_{k})\right),
\end{equation}
where $E_{g}(\xi,\eta)$ is the function of a perfect dispersive element; $\phi_{g}(\xi,\eta,t_{k})$ is the wavefront aberration formed by passing through the dispersive element at the $i$-th epoch; and $FT\{o\}$ represents the Fourier transform of the function of $o$. The function of the dispersive element, $E_{g}(\xi,\eta)$, is a periodic function; hence, the Fourier transform of the function gives different tilts depending on the wavelength to the electric field on the F2 plane. The light of each wavelength is focused to a different detector position. Given that the center position of the densified spectrum at the wavelength of $\lambda(i)$ is $(x_{\lambda}, y_{\lambda})$, the complex amplitude of each densified pupil spectrograph on the detector 1 at the $k$-th epoch is presented as
\begin{equation}
	\label{eqn:E_P4}
	E_{P4}(x,y,t_{k}) = E_{0}(x-x_{\lambda},y-y_{\lambda},\lambda(i))\exp\left(i\frac{2\pi}{\lambda(i)}\phi_{P3}(x-x_{\lambda},y-y_{\lambda},t_{k})\right) * FT\left\{ \exp\left(i\frac{2\pi}{\lambda(i)}\phi_{g}(\xi,\eta,t_{k})\right) \right\}.
\end{equation}
If the wavefront aberration is not generated after passing through the dispersive element (i.e., $\phi_{g}(\xi,\eta,t_{k})=0$), the intensity distribution on detector 1 will be completely independent of the phase error of the incident light; $|E_{P4}(x,y,t_{k})|^{2}=|E_{0}(x-x_{\lambda},y-y_{\lambda},\lambda(i))|^{2}$. Thus, the densified pupil spectrograph can stabilize the spectra of the densified sub-pupils against the incident light phase if the dispersive element is an ideal one. 

However, no dispersive element generates the ideal diffracted wavefront (i.e., $\phi_{g}(\xi,\eta,t_{k}) \neq 0$). When the wavefront error of $\phi_{g}(\xi,\eta,t_{k})$ is smaller than the wavelength of $\lambda(i)$, $\exp\left(i\frac{2\pi}{\lambda(i)}\phi_{g}(\xi,\eta,t_{k})\right) \simeq 1 + i\frac{2 \pi}{\lambda(i)} \phi_{g}(\xi, \eta, t_{k})$. The wavefront error of $\phi_{g}(\xi, \eta, t_{k})$ is written as the sum of ripples with various periods; therefore, the Fourier transform of $\exp\left(i\frac{2\pi}{\lambda(i)}\phi_{g}(\xi,\eta,t_{k})\right)$ does not generate only the main densified sub-pupils at the expected position of $(x=x_{\lambda}, y=y_{\lambda})$ but also faint sub-pupils around the expected position \citep{Traub+2010}. The copied faint sub-pupils are called "speckles" in the field of the high-contrast science. When the ripple period is $\xi_{0}$, the faint speckle is positioned at $\pm \frac{\lambda(i)}{\xi_{0}}$ away from the expected position of $(x=x_{\lambda}, y=y_{\lambda})$. Here, a tilt error in the phase of $\phi_{P3}(x-x_{\lambda}, y-y_{\lambda}, t_{k})$ moves the Airy pattern position on the dispersive element, and the displacement changes the wavefront error generated after passing through the dispersive element. When the Airy disk displacement on the dispersive element caused by the tilt error is small compared to the period of the ripples, the diffracted wavefronts at two epochs, $t=t_{1}$ and $t_{2}$, are related as follows: 
\begin{equation}
	\label{eqn:fai_g}
	\phi_{g}(\xi,\eta,t_{2}) \simeq \phi_{g1} + \frac{\partial}{\partial \xi} \phi_{g1} \Delta \xi +\frac{\partial}{\partial \eta} \phi_{g1} \Delta \eta,
\end{equation}
where $\Delta \xi$ and $\Delta \eta$ are the Airy disk displacements on the dispersive element between the two epochs along the $\xi$ and $\eta$ directions, respectively, and $\phi_{g1}$ represents the diffracted wavefront error at $t=t_{1}$; $\phi_{g1} \equiv \phi_{g}(\xi,\eta,t=t_{1})$. The observed spectra at $t=t_{1}$ and $t_{2}$ are written as follows based on the abovementioned considerations:
\begin{eqnarray}
	\label{eqn:I_4}
	I_{P4}(x,y,\lambda(i),t_{1}) &=& \left| E_{0}(x-x_{\lambda},y-y_{\lambda},\lambda(i)) * FT\left\{ 1 + i\frac{2\pi}{\lambda(i)}\phi_{g1} \right\}\right|^{2} \nonumber \\
	I_{P4}(x,y,t_{2}) &\simeq & \left| E_{0}(x-x_{\lambda},y-y_{\lambda},\lambda(i)) * FT\left\{ 1 + i\frac{2\pi}{\lambda(i)} \left(\phi_{g1} + \frac{\partial}{\partial \xi} \phi_{g1} \Delta \xi + \frac{\partial}{\partial \eta} \phi_{g1} \Delta \eta \right) \right\}\right|^{2}.	
\end{eqnarray}
The difference of the spectra at $t=t_{1}$ and $t_{2}$ is 
\begin{eqnarray}
	\label{eqn:I_4_difference}
	I_{P4}(x,y,\lambda(i),t_{2}) &-& I_{P4}(x,y,\lambda(i),t_{1}) \nonumber \\ 
	&\simeq &  \left| E_{0}(x-x_{\lambda},y-y_{\lambda},\lambda(i)) * FT\left\{i\frac{2\pi}{\lambda(i)} \left(\frac{\partial}{\partial \xi} \phi_{g1} \Delta \xi + \frac{\partial}{\partial \eta} \phi_{g1} \Delta \eta \right) \right\}\right|^{2} \nonumber \\
	&+& \left(E_{0}(x-x_{\lambda},y-y_{\lambda},\lambda(i)) * FT\left\{ i\frac{2\pi}{\lambda(i)} \left(\frac{\partial}{\partial \xi} \phi_{g1} \Delta \xi + \frac{\partial}{\partial \eta} \phi_{g1} \Delta \eta \right) \right\} \right) + c.c.,  
\end{eqnarray}
where $c.c.$ stands for the complex conjugate of the second term in the right-hand side of the equation. Thus, the telescope pointing error induced the change of the recorded intensity under the condition that the dispersive element generates the diffracted wavefront error, and the change of the intensity between the two epochs limits the precision of the radial velocity measurements.  

The speckles at a wavelength affect the main densified sub-pupils at the different wavelengths; thus, the ripples along the spectral direction on the dispersive element have a more significant impact on the precision compared to those along the direction perpendicular to that. For simplicity, we considered how one ripple with a particular period forms the speckle on the detector and limits the precision of the radial velocity measurements. The diffracted wavefront error, $\phi_{g1}$, is written as the sum of the phase and amplitude ripples with a spatial period of $\eta_{0}$ parallel to the dispersion direction:
\begin{equation}
	\label{eqn:phi_g1}
	\phi_{g1} = a \cos \left(2\pi \frac{\eta}{\eta_{0}} + \alpha \right) + i b \cos \left(2\pi \frac{\eta}{\eta_{0}} + \beta \right),
\end{equation}
where $a$ and $b$ are the amplitudes of the ripples in terms of the phase and the amplitude, respectively, and $\alpha$ and $\beta$ are their spatial phases on the dispersive element. From Equations \ref{eqn:I_4_difference} and \ref{eqn:phi_g1}, the difference of the spectra at two epochs can be rewritten as
\begin{equation}
	\label{eqn:I_4_difference_final}
	I_{P4}(x,y,\lambda(i),t_{2}) - I_{P4}(x,y,\lambda(i),t_{1}) \simeq  I_{+} + I_{-},
\end{equation}	
where 
\begin{equation}
	\label{eqn:I_pm}
	I_{\pm} = \Sigma_{i}\left(\frac{2\pi}{\lambda(i)}\right)^{2}\left(\frac{\pi\Delta \eta}{\eta_{0}} \right) \left(\frac{\pi \Delta \eta}{\eta_{0}} \mp 1 \right) \left(a^{2}+b^{2} \pm 2ab\sin(\alpha - \beta)\right) I_{0} \left(x, y - y_{\lambda} \pm\frac{\lambda(i)}{\eta_{0}}, \lambda(i)\right),
\end{equation}
where $I_{0}(x,y-y_{\lambda},\lambda(i))$ is the intensity distribution of the ideal densified sub-pupil at the wavelength of $\lambda(i)$. The diffracted wavefront error produces two asymmetric speckles at $(x,y=y_{\lambda}\pm\frac{\lambda(i)}{\eta_{0}})$, whose intensities are characterized by a combination of $a$, $b$, and $\Delta\eta$. When the spatial frequency of a ripple on the diffraction grating is the diffraction limit along the spectral direction, the shift of the speckle from the original position is equal to the size of each spectral element. As shown in Equation \ref{eqn:I_pm}, the single spectral element is affected by the speckles over the observing wavelength range. Moreover, no pinned speckles formed by the interferences between the rings of the main Airy disk and the speckles exists because the densified sub-pupil is not spread on the detector plane like the diffraction image. 

\begin{figure}
	 \centering
	\includegraphics[scale=0.1,height=8cm,clip]{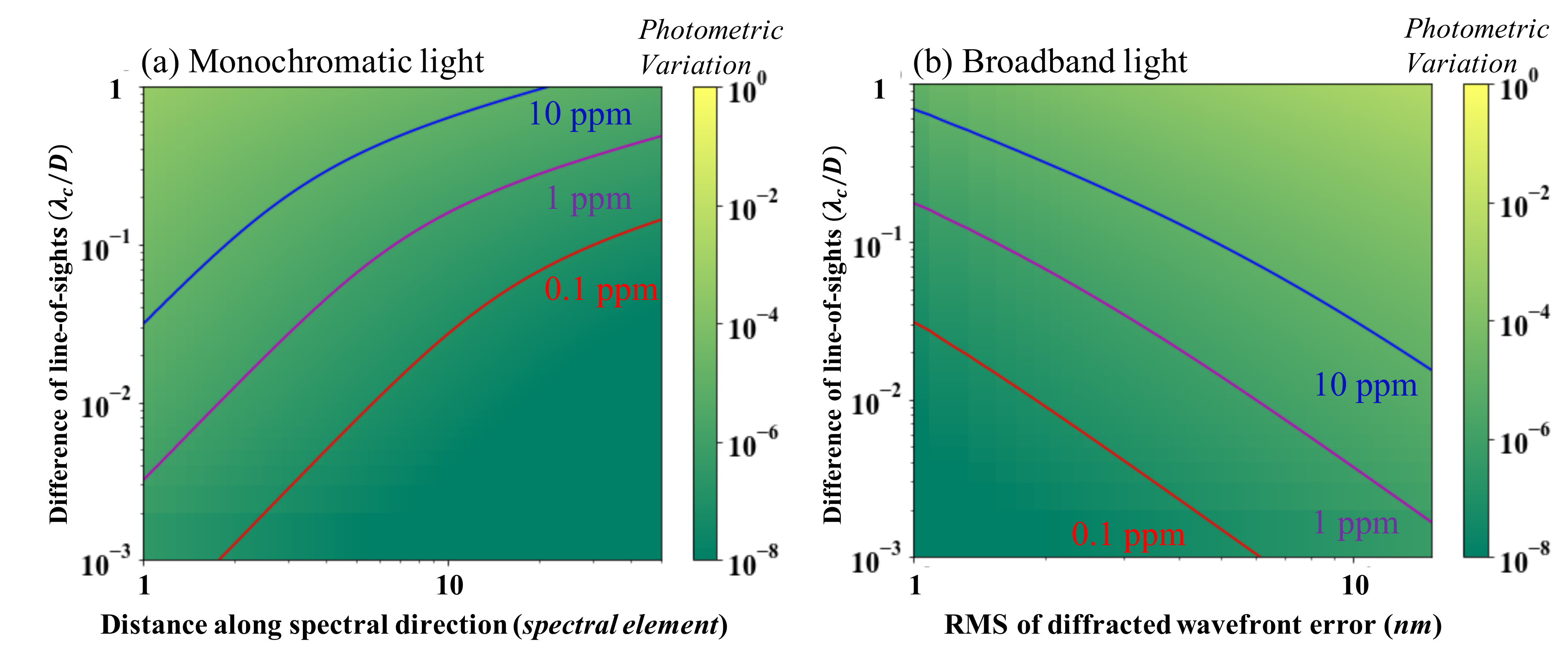}
	\caption{(a) Photometric variation at the wavelength of 500 $nm$ caused by a change in the monochromatic speckle patterns induced by the difference between the telescope line-of-sights at two epochs as a function of the distance along the spectral direction normalized by the spectral element. The root mean square of the diffracted wavefront error was set to approximately 10 $nm$. (b) Photometric variation at the wavelength of 500 $nm$ caused by a change in the speckle patterns of 100 spectral elements around its wavelength as a function of the RMS of the diffracted wavefront error.}
	\label{fig:photometric_wavefront}
\end{figure}

The non-uniformity of the reflectance on the dispersive element is expected to be much smaller than the wavefront error amplitude. While the reflectance non-uniformity could be reduced down to 0.5 $\%$ in visible thanks to a precise control of the coating thickness \citep[e.g.,][]{Cohen+2004, Lightsey+2012}, the phase error of the diffracted wavefront by a reflection grating was approximately 20 - 100 $nm$ peak-to-valley on the scale of 100 $mm$ in that wavelength regime \citep{Lee+2012, Vannoni+2017}. Under a condition that $a >> b$, Equation \ref{eqn:I_pm} approximately becomes
\begin{equation}
	\label{eqn:I_pm_sim}
	I_{\pm} \sim \Sigma_{i}\left(\frac{2\pi a}{\lambda(i)}\right)^{2}\left(\frac{\pi\Delta \eta}{\eta_{0}} \right) \left(\frac{\pi \Delta \eta}{\eta_{0}} \mp 1 \right) I_{0} \left(x, y - y_{\lambda} \pm\frac{\lambda(i)}{\eta_{0}}, \lambda(i)\right).
\end{equation}
Equation \ref{eqn:I_pm_sim} was derived under a simple assumption that there are only the phase and amplitude ripples with a single period of $\eta_{0}$. However the noise floor of the photometry for each spectral element, $\epsilon(i)$, should be determined by the sum of the speckles produced by ripples with various spatial periods. The amplitude distribution of the phase error is determined by the power spectral density (PSD) of the diffracted wavefront error. In this study, the PSD was set as follows based on a simple assumption that the PSD is the same as the structure function of an astronomical telescope mirror:
\begin{equation}
	\label{eqn:PSD}
	PSD (k) =  \frac{A}{1 + \left(\frac{k}{k_{0}}\right)^{3}},
\end{equation}
where $k$ is the spatial frequency. $A$ and $k_{0}$ were set to 400 $nm^{2}cm^{2}$ and 0.2 $cycles/cm$ as a fiducial value, respectively. When the diffraction grating size is 10 $cm$, the root mean square (RMS) of the diffracted wavefront error is approximately 10 $nm$, and the peak-to-valley error at the spatial frequency of 0.1 $cm^{-1}$ is approximately 20 $nm$, almost corresponding to those of the previous study \citep{Vannoni+2017}. The left panel of Figure \ref{fig:photometric_wavefront} shows the photometric variation at 500 $nm$ due to the difference of the monochromatic speckle patterns at two epochs induced by a change in the telescope line-of-sight. The monochromatic speckles are fainter as separating from the original image position; thus, the noise floor is mainly determined by only the speckles of the surrounding spectral elements. 

When the period of the phase ripple is matched to the full width half maximum (FWHM) of the Airy disk formed on the dispersive element, the position of its speckle is shifted to the spectral element next to the original one. The noise floor of each spectral element can be easily calculated by normalizing the spatial period of the ripples by the FWHM of the Airy disk. Given that the phase ripple amplitude with the normalized period of $\eta_{m}$ is $a_{m}$, Equation \ref{eqn:I_pm_sim} can be rewritten as follows:
\begin{equation}
	\label{eqn:I_pm_2}
	I_{\pm} = \Sigma_{i} \Sigma_{m} \left(\frac{2\pi a_{m}}{\lambda(i)}\right)^{2}\left(\frac{\pi\Delta \eta}{\eta_{m}} \right) \left(\frac{\pi \Delta \eta}{\eta_{m}} \mp 1 \right) I_{0} \left(x, y - y_{\lambda} \pm\frac{\lambda(i)}{\eta_{m}}, \lambda(i)\right).
\end{equation}
The right panel of Figure \ref{fig:photometric_wavefront} shows the photometric precision limited by the difference between the speckle patterns of 100 adjacent spectral elements at two epochs. To achieve the photometric precision of 1 $ppm$, the offset of the telescope line-of-sights at two epochs should be reduced to 0.03 $\frac{\lambda}{D}$ in the fiducial case that the RMS of the wavefront error is 10 $nm$. Thus, a combination of the quality of the diffracted wavefront and the offset of the telescope line-of-sight at two epochs determines the difference of the spectra at two epochs, $I_{P4}(x,y,\lambda(i),t_{2})-I_{P4}(x,y,\lambda(i),t_{1})$, corresponding to the noise floor of the photometry for each spectral element, $\epsilon(i)$. The limitation on the precision of the radial velocity measurements was estimated through Equation \ref{eqn:limitation_system_latter_case}. In Section \ref{sec:evaluation}, we estimate the precision of the radial velocity measurements under an appropriate observing condition.


\subsection{Monitor of the long-term drift} \label{subsec:sensor_drift}
\begin{figure}
	 \centering
	\includegraphics[scale=0.1,height=12.8cm,clip]{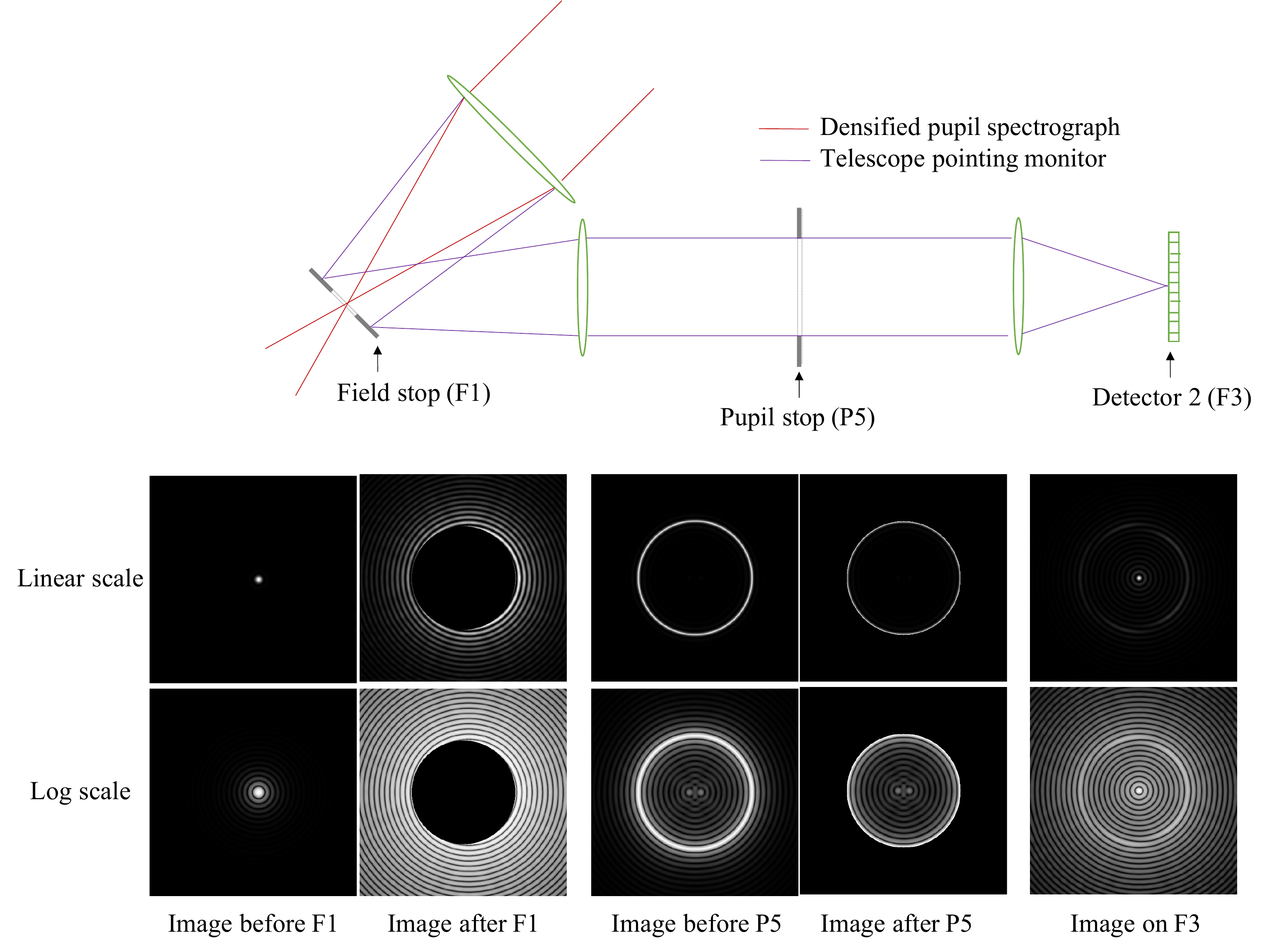}
	\caption{Conceptual diagram of a telescope line-of-sight monitor. The light was divided into two by the field stop. The light inside the field stop is introduced to the densified pupil spectrograph indicated by the red line. The remaining light was reflected by the field stop and introduced to the monitor, as indicated by the purple line. The upper and lower images show the intensity distributions on each plane on linear and log scales, respectively. The simulation images were generated under the following conditions; 1) a clear circular aperture with a tilt error of 1 $\frac{\lambda}{D}$ along the horizontal direction prepared as the entrance pupil; 2) a field stop radius set to 10 $\frac{\lambda}{D}$; 3) a pupil stop diameter equal to that of the entrance pupil.}
	\label{fig:concept_pointing_monitor}
\end{figure}

The precision of the radial velocity measurements is ultimately limited by the difference of the telescope line-of-sights at two epochs. In exchange for the high stability against the telescope pointing errors, the tilt error cannot be estimated from the spectra recorded by the densified pupil spectrograph because the spectra are insensitive to the phase of the pupil plane. It is difficult even for conventional spectrographs to precisely measure the telescope line-of-sight because the light from an astronomical object is dispersed along the wavelength direction on the detector. We propose herein a sensor working  not only for the densified pupil spectrograph but also for conventional ones. Figure \ref{fig:concept_pointing_monitor} shows the conceptual design of the telescope line-of-sight monitor. After the light outside the field stop is introduced to the telescope line-of-sight monitor, the light is collimated and passed through a pupil stop, and the light is then focused on a detector. The key parameter of this monitor is the pupil stop diameter. When the diameter exactly matches that of the entrance pupil (i.e., the primary mirror), this monitor can work as a telescope line-of-sight monitor. The intensity distribution formed on detector 2 includes an information on the telescope line-of-sight. 

We will mathematically explain the wavefront propagation through the line-of-sight monitor. The complex amplitude formed on the field stop (F1) is the Fourier transform of the electric field on the entrance pupil (P1); $E_{F1}(\xi,\eta,\lambda(i))=FT\{E_{P1}(x,y,\lambda(i))\}$. Based on Babinet's principle, the complex amplitude after the field stop can be written as
\begin{equation}
	\label{eqn:E_F1}
	E_{F1}(\xi,\eta,\lambda(i)) = FT\{E_{P1}(x,y,\lambda(i))\} \left(1 - A_{field-stop}(\xi,\eta)\right),
\end{equation}
where $A_{field-stop}(\xi,\eta)$ is the aperture function of the field stop assumed to be a clear circular aperture. The complex field on the pupil stop (P5) is 
\begin{equation}
	\label{eqn:E_P5}
	E_{P5}(x,y,\lambda(i)) = E_{P1}(x,y,\lambda(i)) - E_{P1}(x,y,\lambda(i)) * FT\{A_{field-stop}(\xi,\eta)\},
\end{equation}
where $FT\{A_{field-stop}(\xi,\eta)\}$ is the Bessel function of the first kind of order one. Most of the light is canceled by the destructive interference, and the remaining faint light is concentrated at the edge of the pupil stop. After passing through the pupil stop with the aperture function of $A_{pupil-stop}(x,y)$, the complex amplitude on detector 2, $E_{F3}(\xi,\eta)$, is mathematically derived as follows:
\begin{equation}
	\label{eqn:E_F3}
	E_{F3}(\xi,\eta,\lambda(i)) = \left[FT\{E_{P1}(x,y,\lambda(i))\} - FT\{E_{P1}(x,y,\lambda(i))\} A_{field-stop}(\xi,\eta)\right] * FT\{A_{pupil-stop}(x,y)\}.
\end{equation}
When the pupil stop diameter is equal to that of the entrance pupil, $FT\{A_{pupil-stop}(x,y)\}$ becomes the ideal ASF of the telescope. The electric field on detector 2, $E_{F3}(\xi,\eta,\lambda(i))$, is formed through a convolution of the image on the field stop with the Fourier transformation of the pupil stop. Concentric circle and ring patterns with the same period as that of the ASF base on the field stop are formed on the detector. The simulated image on detector 2 (Figure \ref{fig:concept_pointing_monitor}) is different from the original PSF. The amplitude of $E_{F3}(\xi,\eta)$ is maximized when the center position of the ASF is matched to that of the complex amplitude in the parentheses of Equation \ref{eqn:E_F3}. The telescope line-of-sight was reflected in the center position of the amplitude in the parentheses of Equation \ref{eqn:E_F3}; thus, this monitor allowed us to measure, in principle, the pointing error from the intensity distribution formed on the detector. 

\begin{figure}
	 \centering
	\includegraphics[scale=0.1,height=11cm,clip]{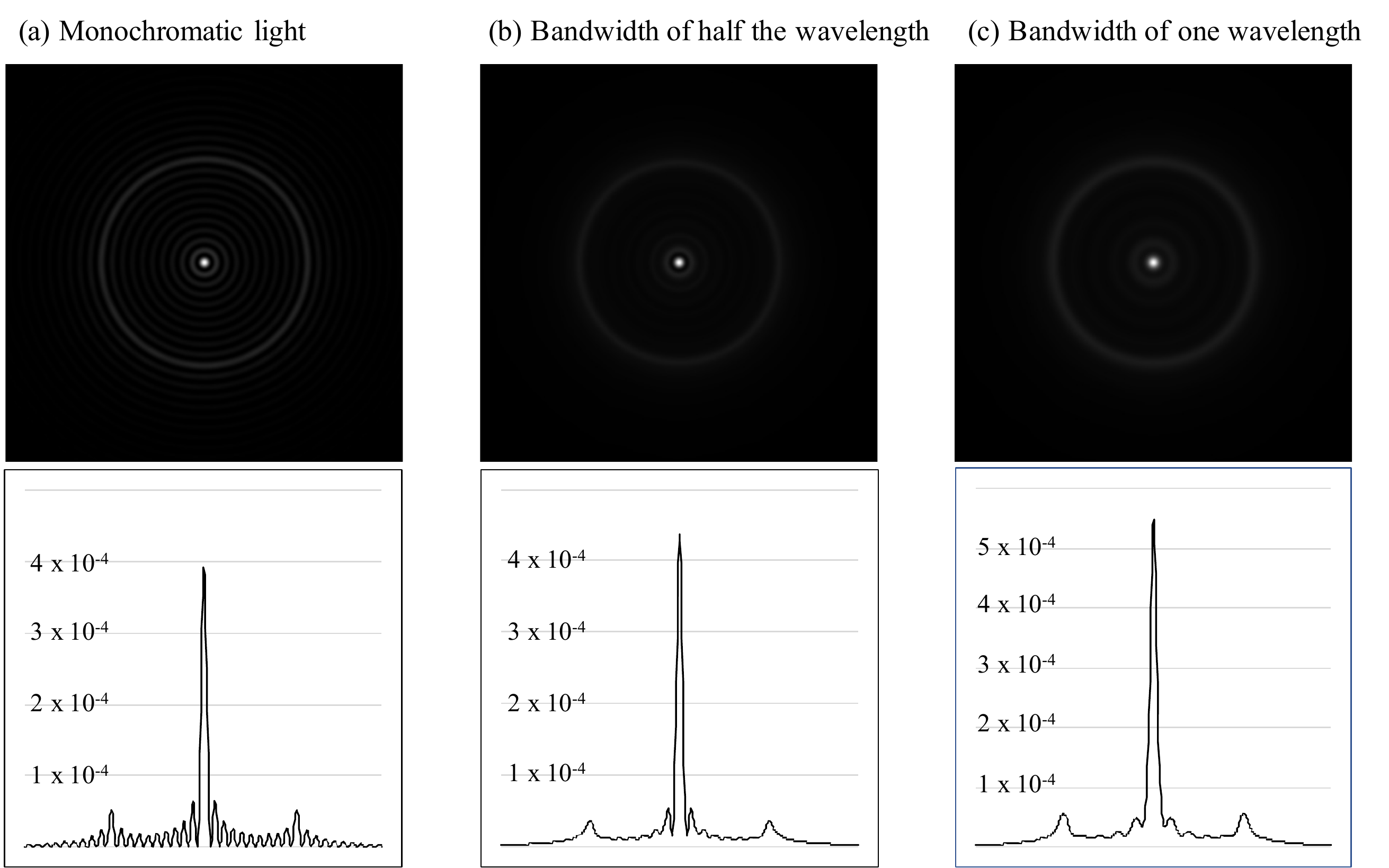}
	\caption{Images formed on the telescope line-of-sight monitor for (a) the monochromatic light, (b) the bandwidth of half the central wavelength, and (c) the bandwidth equivalent to the central wavelength. The upper panels show the two-dimensional intensities on the linear scale. The lower panels show their intensity distributions along the horizontal direction. The values of the vertical axis represent the average ratio of the intensity on the sensor to the peak intensity of the original PSF. The tilt of the incident light was set to 0.}
	\label{fig:bandwidth}
\end{figure}

This sensor works in the broadband light because there are no chromatic optical elements in the sensor system. Figure \ref{fig:bandwidth} shows the intensity images formed on detector 2 for three different bandwidths. The image gradually blurred as the bandwidth increased because the core size was proportional to the wavelength. However, the center core of the image measured for estimating the telescope line-of-sight was still clear even under the condition of the bandwidth being equal to the observing wavelength. Note that the collimator and the camera systems should be designed such that they do not generate a chromatic aberration. 

Even though the bandwidth can be increased, most of the light from an astronomical object is required to be introduced to the densified pupil spectrograph. The pointing error must be measured with the remaining faint light. As shown in the right panel of Figure \ref{fig:motion_loss}, the number of photons introduced to the monitor was more limited as the field-stop radius increased. For example, the peak intensity of the core formed on detector 2 was reduced by factors of approximately $1.5 \times 10^{-3}$ and $4 \times 10^{-4}$ for the field-stop radii of 5 and 10 $\frac{\lambda}{D}$, respectively. According to the previous study of \cite{Sandrine+2004}, the measurement accuracy of the centroid position was equal to the ratio of its FWHM to the square root of the number of the photons for the photon-noise-limited case. The FWHM of the centroid corresponded to the angular resolution of the telescope; therefore, the precision of the telescope line-of-sight monitor is written as
\begin{equation}
	\sigma_{sensor} = \frac{\theta_{angular}}{\sqrt{N_{sensor}}},
\end{equation}
where $\theta_{angular}$ is the angular resolution of the telescope, and $N_{sensor}$ is the number of photons introduced to this monitor, which is determined by a combination of the number of the photons in the incident beam to the telescope and the field-stop size. 

\begin{figure}
	 \centering
	\includegraphics[scale=0.1,height=8cm,clip]{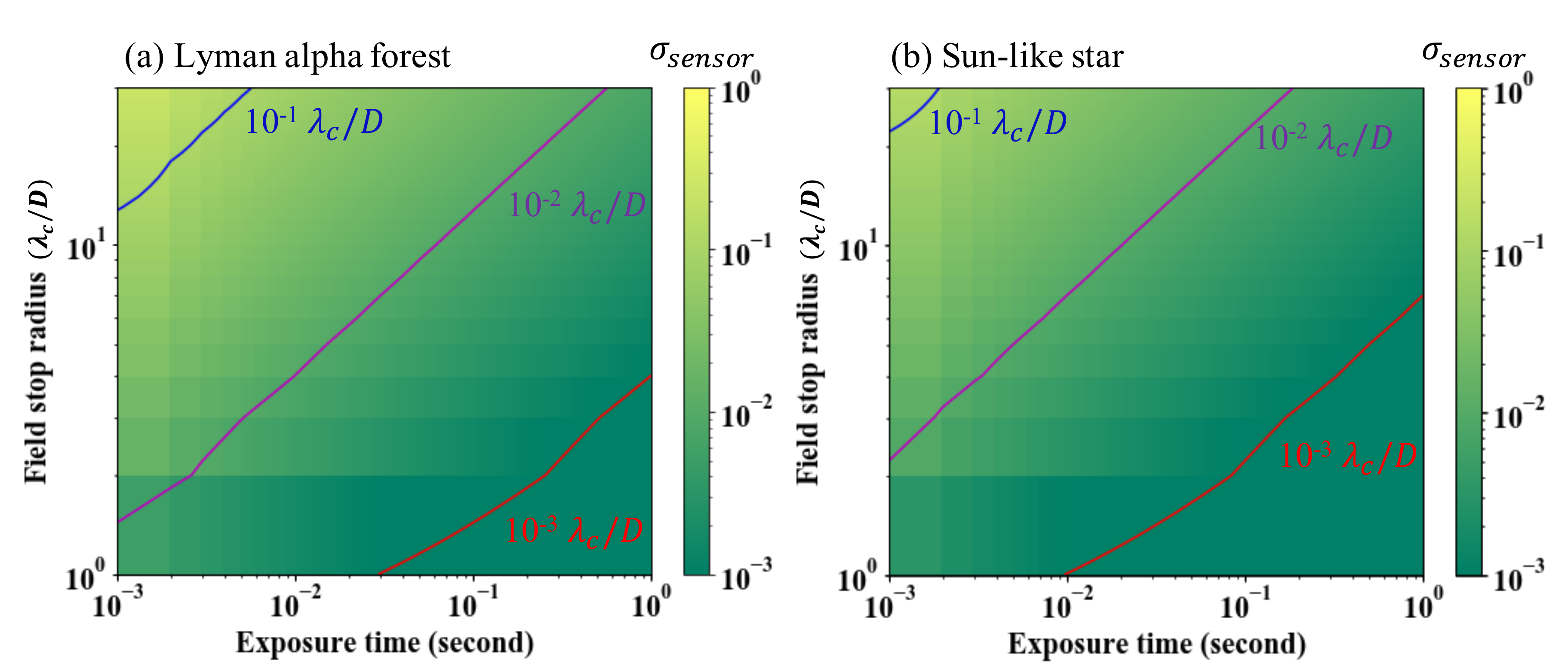}
	\caption{Precision of the telescope line-of-sight monitor for the (a) Lyman alpha forest of a QSO and (b) the spectrum of a Sun-like star at 30 $pc$ as a function of the field stop radius and exposure time. The blue, purple, and red lines represent the precisions of $10^{-1}$, $10^{-2}$, and $10^{-3}$ $\frac{\lambda_{c}}{D}$, respectively. The monitor bandwidth was simply assumed to be equal to the central wavelength of 0.5 $\micron$. The telescope diameters for (a) and (b) were set to 15 $m$ and 2 $m$, respectively. The optical throughput, except for the losses caused by the field stop and the pupil stop, was set to 0.9. The flux of the Lyman alpha spectrum was simply set to 0.23 $m$Jy, corresponding to 18.0 mag in the $V$ band. The flux of the Sun-like star was calculated based on the parameters listed in Table \ref{tab:target_and_telescope}.}
	\label{fig:precision_sensor}
\end{figure}

Figure \ref{fig:precision_sensor} shows the achievable precision of the telescope line-of-sight monitor for a Lyman alpha forest of a QSO and a nearby Sun-like star when observing using 15 $m$- and 2 $m$-diameter telescopes, respectively. When observing a Sun-like star at 30 $pc$ with the 2 $m$ class telescope, this monitor collects a sufficient number of photons for only a few milli-second under the condition of the field-stop radius being less than 5 $\frac{\lambda}{D}$. The line-of-sight jitter is possibly monitored with an accuracy of 0.01 $\frac{\lambda}{D}$ in the same manner as the wavefront sensors for adaptive optics \citep[e.g.,][]{Hardy+1998,Guyon+2005} and the low-order wavefront sensors for a coronagraph \citep{Guyon+2009}. In contrast, the slow telescope line-of-sight with a timescale of a few tens of seconds can be measured with this monitor for the Lyman alpha forest because the QSO is much fainter than the nearby stars. The required exposure time strongly depends on the telescope diameter and field-stop diameters. The telescope diameter is smaller; thus, this monitor measures only the slower change in the telescope line-of-sight. However, the high-frequency line-of-sight jitter is randomly changed and has no significant impact on the radial velocity measurements. The signal variation caused by the high-frequency jitter could be smoothed out by averaging the data over a long integration time. In other words, the long-term drift will increase the difference of the telescope line-of-sights at two epochs and limit the precision of the radial velocity measurements. As shown in Figure \ref{fig:precision_sensor}, this monitor can fully measure the long-term drift; hence, the precision of the radial velocity measurements could be improved by working at the same time when the densified pupil spectra are recorded. The telescope line-of-sight monitor is suitable for the radial velocity measurements of distant QSOs and nearby stars. The appropriate field-stop diameter should be determined by the balance between the signal fluctuations caused by the motion loss and the shot noise of an astronomical object. If the telescope pointing jiitter is small, the field-stop size could be reduced, thanks to the smaller motion loss. We will determine the appropriate field-stop size under an appropriate condition in Section \ref{sec:evaluation}.  

Finally, we mention the difference between this monitor and the Lyot coronagraph \citep{Lyot+1939}. The monitor configuration is the same as that of the Lyot coronagraph. The complex amplitude shown in Equation \ref{eqn:E_F3} is equal to that formed on the detector plane of the Lyot coronagraph. The only difference between the two systems is the pupil/Lyot stop diameter. The Lyot stop diameter should be smaller than that of the entrance pupil because the unwanted diffracted light from the edge of the occulting mask/field stop on the focal plane is blocked. In contrast, the pupil stop of this monitor transmits half of the light coming from the field stop. The simulated images before and after the pupil stop shown in Figure \ref{fig:concept_pointing_monitor} depict the outer half of the bright ring in the image before P5 is blocked by the pupil stop.

\section{Direct measurement of the universe's expansion history} \label{sec:evaluation}
Section \ref{sec:concept} described an approach for enabling high-precision radial velocity measurements through developing the densified pupil spectrograph originally proposed for the mid-infrared transit spectroscopy of exoplanets. This instrument concept was composed of a high-dispersion densified pupil spectrograph and a monitor for measuring the telescope line-of-sight, working at the same time when the densified pupil spectra are recorded. The present section presents a new compact and simple optical design developed from the existing densified pupil spectrograph. This new design allows us to simultaneously acquire the spectra of approximately 10 astronomical objects at different field-of-views. We evaluate how much the expansion history of the Universe at high redshits could be directly measured with this instrument design under an appropriate assumption.

\subsection{Spectrograph design for direct measurement of the Universe's expansion history} \label{subsec:instrument_design}
This subsection shows an instrument design optimized for the direct measurement of the Universe's expansion history. We first determine the main optical parameters of the spectrograph based on the performance of the radial velocity instrument analytically derived in the previous sections. We next present the optical design of a spectrograph satisfying the main optical parameters.

\subsubsection{Specification} \label{subsubsec:specification}
We determined the main parameters of the spectrograph, such that the two main sciences can be performed. The optical system introduced herein is a tentative solution. The optical design should be finally optimized under a balance between its spectrograph volume and its specification for enabling the science case. As discussed in Section \ref{sec:basics}, the resolving power of the spectrograph enhanced high $Q_{photon}$ and $Q_{sys}$ factors and reduced an imapct of the systematic instrumental noise on the precision. As shown in Figure \ref{fig:Q_factor}, both the $Q_{photon}$ and $Q_{sys}$ factors for the Lyman alpha forest were almost constant in the regime of the resolving power higher than 10,000. Considering that a higher resolving power limits the wavelength range more and increases the spectrograph volume, we set the resolving power to 10,000.

We focused on obtaining the spectra of the QSOs at z $>$ 2 because it was difficult to present a compact optical design in the ultraviolet wavelength regime. The number of lens materials available for that regime was very limited, and a chromatic aberration remained. The longer wavelength edge of the Lyman alpha forest is 365 $nm$ for the QSOs at $z = 2$; thus, it was prefarable to set the shorter edge of the wavelength range to that number. The Lyman alpha forests of the high-redshift QSOs at z $\sim$ 5 can be fully obtained when the longer edge of the wavelength range is 670 $nm$. Thus, we set the wavelength range of the spectrograph to $<$ 365 to 670 $nm$.

We determined the optimum angular size of the field stop for the field-of-view of the spectrograph. As discussed in Section \ref{sec:concept}, the field-stop radius should be determined by the balance between the photometric variation caused by the motion loss and the precision of the telescope line-of-sight monitor. The photometric variation due to the motion loss induced by the telescope line-of-sight jitter decreased as the field-stop radius became larger (Figures \ref{fig:motion_loss}). In other words, the line-of-sight jitter requirement was stricter for the smaller field-stop radius. For example, to reduce the photometric variation caused by the motion loss down to 10 $ppm$, the field-of-view should be larger than 4 $\frac{\lambda}{D}$ under the condition of the line-of-sight jitter being 0.05 $\frac{\lambda}{D}$. In contrast, the amount of the light introduced to the telescope line-of-sight monitor decreased for the larger field stop radius, and the monitor precision was limited. As shown in Figure \ref{fig:precision_sensor}, a long exposure time was required for achieving a higher monitor precision because the Lyman alpha forest stamped on the spectrum of a high-redshift QSO was very faint. The instrumental noise could be less than 1 $ppm$ when the difference of the line-of-sights at two epochs is reduced to 0.001 $\frac{\lambda}{D}$ (Figure \ref{fig:photometric_wavefront}). The field stop radius should be less than 4 $\frac{\lambda}{D}$ to achieve the measurement precision of 1 $ppm$ with a modest exposure time of less than 1,000 s. Thus, the field-of-view was set to 4 $\frac{\lambda}{D}$. 

\subsubsection{Optical design} \label{subsubsec:optical_design}
\begin{figure}
	 \centering
	\includegraphics[scale=0.1,height=16cm,clip]{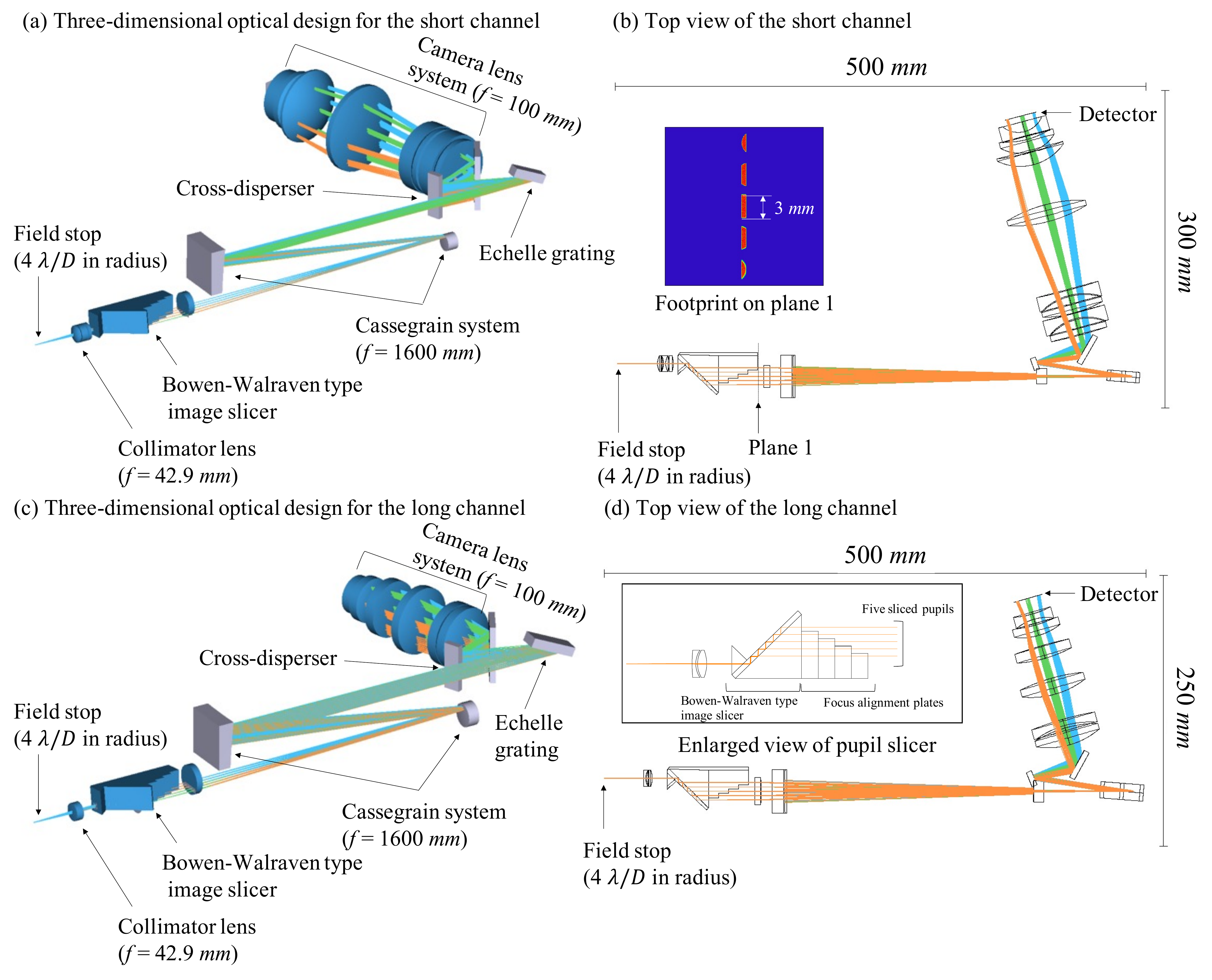}
	\caption{High-dispersion spectrograph design optimized for the (a)(b) short (305 - 500 $nm$) and (c)(d) long (463 - 660 $nm$) channels. A dichroic mirror between the field stop and the collimator lens, which is not shown in this diagram, divides a beam into two in terms of the wavelength. The color of the ray indicates a different grating order: 29, 23, and 18 grating orders for the blue, green, and orange rays in the short channel and 35, 29, and 25 grating orders in the long one, respectively. Each beam has a field of view of 4 $\frac{\lambda}{D}$ in radius. The entrance beam with 3 $mm$ diameter was divided into five without any densification. The footprint on the plane 1 after the Bowen-Walraven type image slicer was used is shown in the left upper panel of (b).}
	\label{fig:High_dispersion_DPS_design}
\end{figure}

\begin{figure}
	 \centering
	\includegraphics[scale=0.1,height=14cm,clip]{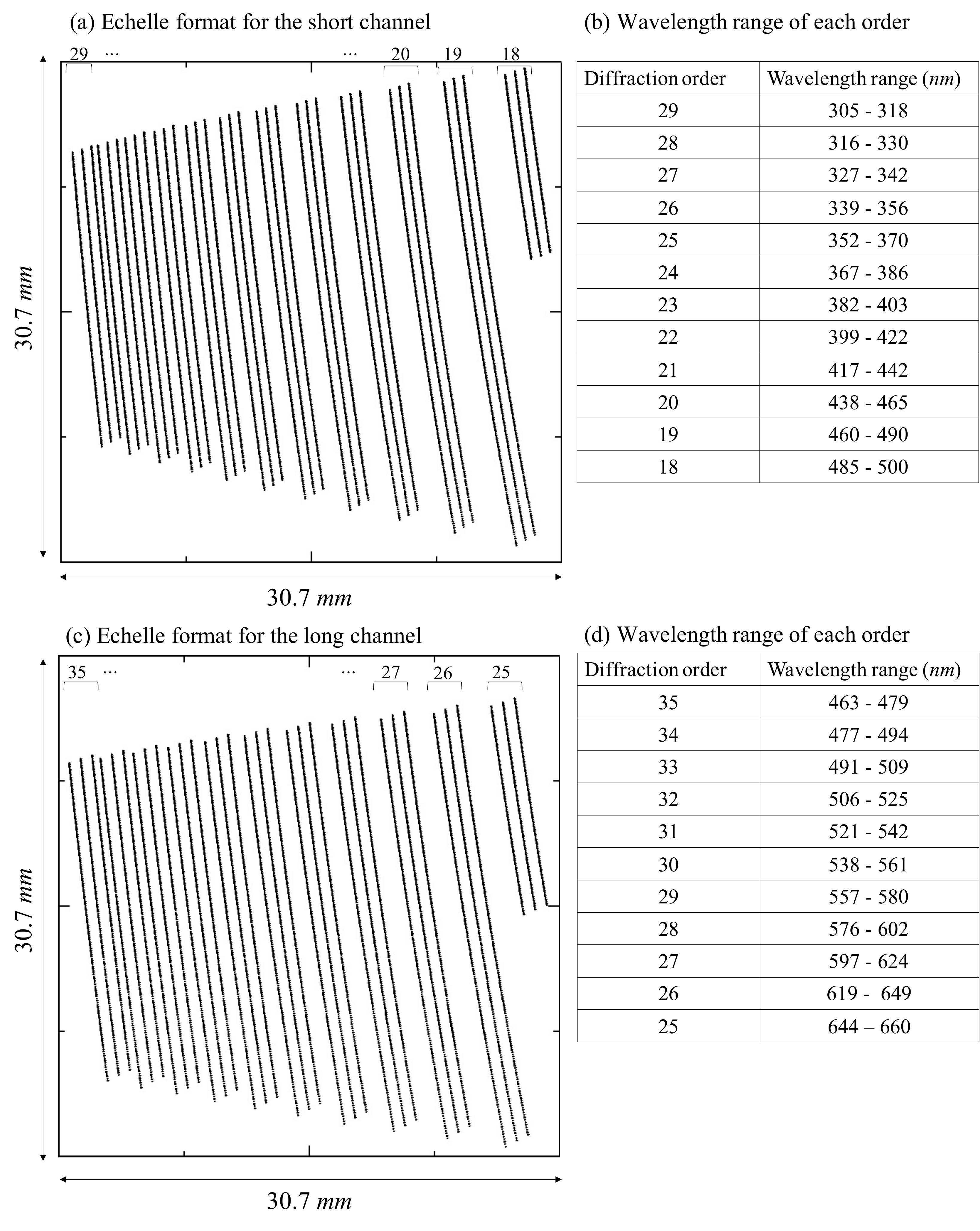}
	\caption{(a)(c) Echelle formats of the short and long channels. The vertical lines indicate the different diffraction orders of 18 to 29 and 25 to 35 for the short and long channels, respectively. The detector size of each channel was assumed to be 30.7 x 30.7 $mm$. The panels (b) and (d) show the wavelength ranges of each order for the short and long channels, respectively.}
	\label{fig:footprints}
\end{figure}

We constructed a new simple and compact optical design based on the determined spectrograph specification. This new design was developed from the existing densified pupil design for the Origins Space Telescope \citep{Matsuo+2018}. Figure \ref{fig:High_dispersion_DPS_design} shows two optical designs optimized for different wavelength ranges. After passing through the field stop with a radius of 4 $\frac{\lambda}{D}$, the beam was first divided into two by a dichroic mirror in terms of the wavelength range: 1) 305 - 500 $nm$ for the short channel; and 2) 463 - 660 $nm$ for the long one. The two designs were almost identical, except for the optical coating, echelle grating, and lens design. After passing through the dichroic mirror, the incident beam with 3 $mm$ diameter was collimated by a lens with a 42.9 $mm$ focal length and divided into five with a compact beam slicer (Bowen-Walraven type image slicer) \citep{Bowen+1938, Walraven+1972}. Although this type of image slicer is widely applied to high-resolution spectrographs \citep[e.g.,][]{Schwab+2011,Tajitsu+2012}, this was the first time it was used as a pupil slicer. The upper panel of Figure \ref{fig:footprints} shows the footprints on the plane after the Bowen-Walraven type image slicer was used. A focus alignment plate was added to the Bowen-Walraven type image slicer. All sub-pupils were focused on the reflection grating without defocus. This new optical design did not require a pupil densification with two concave mirrors because of the small diameter of the entrance beam reduced by the image slicer. The divided beams were collimated by a Cassegrain system composed of two mirrors with an effective focal length of 1800 $mm$. An Echelle grating with a pitch interval of 200 (110) $lines/mm$ for the short (long) channel dispersed the collimated five beams. A cross-disperser of a reflection grating separated the overlapping spectral bands with diffraction orders of 18 to 29 (25 to 35) in the short (long) channel. The Litrrow configuration was chosen. The incident beam angle to the Echelle grating was equal to its brazed angle. The camera lens system was designed under the condition of each spectral element being sampled by approximately 2.5 pixels along the spectral direction at the central wavelength of each order. In this case, the number of each spectral element was approximately 122, which will reduce the impacts of an uncorrected flat-field error and a finite pixel sampling of the PSF on the precision of the radial velocity measurements. Considering that a photon counting Electron Multiplying CCD (EMCCD) camera with a large format of 4k x 4k has now been tested \citep{Daigle+2018}, the format and the pixel pitch of the detector for this spectrograph were set to 2k x 2k and 12 $\micron$, respectively. The left and right panels of Figure \ref{fig:footprints} show the Echelle format of the shorter and longer channels, respectively. The field-of-view of the spectrograph was set to 4 $\frac{\lambda}{D}$ at the longest wavelength of 660 $nm$. The wavelength range of each unit is limited; thus, the anti-reflection coating can be optimized. The optical throughput of the spectrograph was approximately 0.63 assuming that the throughput of each anti-reflection coating was 0.995, the diffraction efficiency was 0.8, and the efficiency of the pupil slicer was 0.9. Considering that the telescope throughput was 0.9 and the quantum efficiency of the detector was 0.8, we set the total throughput to 0.45. The spectrograph size optimized for each band was approximately 550 (L) x 300 (W) x 100 (H) $mm$. Table \ref{tab:optical_parameters} compiles the optical design parameters. 


\begin{table}[htb]
	\begin{center}
		\caption{Parameters of the optical designs optimized for the short and long channels}
  		\begin{tabular}{| l | c | c |} \hline 
    		& Short channel & Long channel \\ \hline \hline
    		Wavelength range & 305 - 500 $nm$ & 463 - 660 $nm$ \\ \hline
    		Resolving power at the central wavelength of each order & 10,800 & 10,800 \\ \hline 
    		Field-of-view & 4 $\frac{\lambda}{D}$ at 500 $nm$ & 4 $\frac{\lambda}{D}$ at 660 $nm$\\ \hline     		
			Incident beam diameter & 3 $mm$ & 3 $mm$ \\ \hline    		
    		Focal length of the collimator lens & 42.9$mm$  & 42.9 $mm$ \\ \hline
    		Number of sub-pupils & 5  & 5 \\ \hline
    		Focal length of the Cassegrain system & 1800 $mm$ & 1800 $mm$ \\ \hline
    		Type of cross-disperser & Reflection grating & Reflection grating \\ \hline
    		Pitch interval of the Echelle grating & 200 $lines/mm$ & 110 $lines/mm$ \\ \hline
    		Brazed angle & 65.5 $degree$ & 65.5 $degree$ \\ \hline
    		Diffraction order & 18 - 29 & 25 - 35 \\ \hline
    		Focal length of the camera lens system & 100 $mm$ & 100 $mm$ \\ \hline
    		Number of samplings along spectral direction & 2.5 & 2.5 \\ \hline
    		Detector format & 2 k x 2 k & 2 k x 2 k \\ \hline
    		Pixel size of the detector & 15 $\micron$ & 15 $\micron$ \\ \hline
  \end{tabular}
  \label{tab:optical_parameters}
  \end{center}
\end{table}




\subsection{Performance} \label{subsec:results}
Based on the specification of the designed spectrograph unit, we evaluated the precision of the radial velocity measurements for direct measurement of the Universe's expansion history. The spectrum of the Lyman alpha forest was generated through the \texttt{$fake\_spectra$} package \citep{Bird:2014a} (Appendix \ref{sec:appendix}). For comparison, the two redshifts of $z=3$ and 4 were focused in this evaluation. Based on the Million Quasars (MILLIQUAS) catalog \citep{Flesch+2021}, the apparent $B$ band mangnitudes of the brightest targets at approximately $z$ = 3 and 4 are 16.01 and 17.29, respectively. We note that both of them are confirmed as QSOs in the catalog. The applied values almost correspond to those of the objects introduced in the previous study of \cite{Liske+2008}.

The telescope for the direct measurement of the Universe's expansion history was assumed to be the LUVOIR-A concept \citep{LUVOIR+2019}. The telescope line-of-sight jitter of the LUVOIR-A concept was set to 0.3 $milli-arcsecond$ \citep{Sacks+2018}. The jitter value almost corresponded to 0.05 $\frac{\lambda}{D}$ at the longest wavelength of 660 $nm$. Considering that the field-stop radius was $4 \frac{\lambda}{D}$ at that wavelength, the motion loss on the field stop caused by the line-of-sight jitter will cause the photometric variation of approximately 10 $ppm$ (Figure \ref{fig:motion_loss}). However, the photometric variation is randomly distributed; thus, the impact of the motion loss on the photometry can be canceled out by adding a number of exposures. The photometric degradation caused by the motion loss was not considered in this evaluation. When we observed a Lyman alpha forest spectrum with an apparent magnitude of 17.3 $mag$ in the $B$ band with a LUVOIR-A concept telescope, the number of the obtained electrons per second for each spectral element was approximately 62 $e/s$ on average. Thanks to the advanced technology for the low-flux detectors in visible \citep[e.g.,][]{Wilkins+2014}, we simply neglected the detector noise (e.g., read noise and dark current).

The precision of the radial velocity measurements was mainly limited by the instrumental systematic noise, wavelength calibration source, and shot noise. The systematic noise floor caused by the difference between the line-of-sights at two epochs can be reduced down to 1 $ppm$, thanks to the line-of-sight sensor. However, the photometric precision was not only limited by the optical wavefront errors, but also by the detector system. For example, the previous studies showed that the photometric precision was an order of 10 $ppm$ based on the characterization of only the detector system in the experiment room \citep{Matsuo+2019, Schlawin+2021}. In this calculation, the systematic noise was assumed to be determined by the root sum square of the two independent noises. The wavelength calibration error was also considered, which is mainly caused by a long-term change in the LFC spectrum, an imperfection of the flat-field light source, and a variation in the signal-to-noise ratio of the calibration source along the wavelength \citep[e.g.,][]{Blackman+2020}. The long-term stability of the source limited the precision of the radial velocity measurements to $\sim$ 2 $cm/s$ \citep{Halverson+2016}. In contrast, the number of the detector samplings for each spectral element in this spectrograph system was approximately 120, which is much larger than those of other high-precision radial velocity instruments. The impacts of the uncorrected flat-field error and the finite pixel samplings of the PSF will be largely mitigated. In this calculation, the uncertainty of the wavelength calibration was simply set to 2 $cm/s$ considering that only the calibration source stability limits the precision. Finally, the precision will fundamentally be limited by the shot noise associated with the spectrum recorded by the densified pupil spectrograph. Table \ref{tab:target_and_telescope} presents all of the parameters used for this evaluation. Note that the impact of a change in the spectrograph temperature on the precision was not considered because the thermal stability of the LUVOIR telescope concept could be achieved to 1 $mK$ \citep{Yang+2019}, which was much smaller than those of the ground-based observatories \citep[e.g.,][]{Blackman+2020}. 

\begin{figure}
	 \centering
	\includegraphics[scale=0.1,height=7cm,clip]{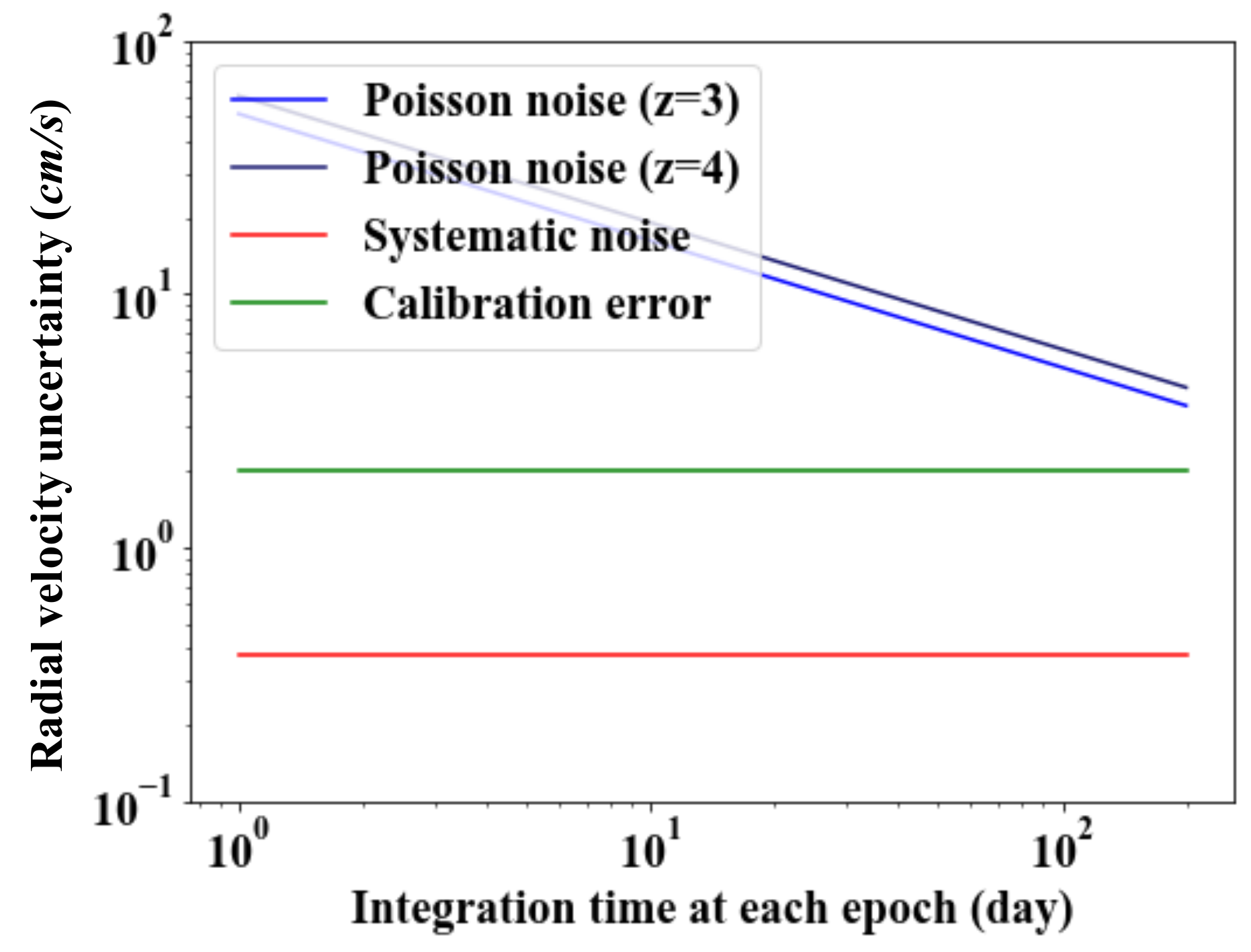}
	\caption{Precision of the radial velocity measurements for the direct measurement of the Universe's expansion history, as a function of the integration time at each epoch. The Lyman alpha forests of QSOs at $z = 3$ and 4 were used (Appendix \ref{sec:appendix}). The parameters used for this evaluation are compiled in Table \ref{tab:parameters_of_target_system}. The wavelength ranges of the Lyman alpha forests were equally divided into 3 to trace the Universe's expansion history at high redshifts. The divided wavelength intervals for the QSOs at $z$=3 and 4 were 603 and 1008 $nm$, respectively.}
	\label{fig:uncertainty_two_science_cases}
\end{figure}

Figure \ref{fig:uncertainty_two_science_cases} shows the uncertainty of the radial velocity measurements for the direct measurement of the Hubble parameters at various redshifts. This proposed instrument concept could reach precision of $\sim$ 3 and 4 $cm/s$ for the Lyman alpha forests of QSOs at $z$ = 3 and 4, respectively, when the integration time at each epoch was 200 days. To measure the expansion history at various redshifts, the wavelength range was equally divided into 3 for both the QSOs at $z$ = 3 and 4. The divided wavelength ranges for the QSOs at $z$ = 3 and 4 were fixed to 603 and 1008 $nm$, respectively. The calibration error will principally limit the precision of the radial velocity measurements when the integration time is infinite. Finally, the systematic error of an order of 10 $ppm$ did not affect the precision for this science case.

\begin{table}[htb]
	\begin{center}
		\caption{Parameters of the targets and the observatories for the direct measurement of the Hubble parameters}
  		\begin{tabular}{| l | c |} \hline 
   			Model spectrum & \texttt{$fake\_spectra$} \\ \hline
   			Flux densities of the brightest QSOs & 0.90 $m$Jy at $z$=3, 0.36 $m$Jy at $z=$4 \\ \hline
   			Flux densities of the faint QSOs & 0.23 $m$Jy at $z=$3, 0.091 $m$Jy at $z$=4 \\ \hline
    		Telescope diameter & 1500 $cm$ \\ \hline
			Line-of-sight jitter & 0.3 $mas$ \\ \hline 
			Number of the field-of-views & 10 \\ \hline 			
			Systematic noise of the optical system & 1 $ppm$ \\ \hline    		
   		 	Systematic noise of the detector system & 10 $ppm$ \\ \hline
   			Wavefront calibration error & 2 $cm/s$  \\ \hline
  \end{tabular}
  \label{tab:target_and_telescope}
  \end{center}
\end{table}

\subsection{Expansion history of the Universe} \label{subsec:expansion}

The Lyman alpha forest of the QSO is formed by the absorptions of neutral hydrogen atmos between the QSO and the observer; hence, the redshift drifts of the hydrogen medium at various redshifts can be directly measured from the difference between the radial velocity signals at two epochs with a time span of more than a few years \citep{Sandage+1962, Loeb+1998}. In this subsection, we introduce how the instrument concept could constrain the Universe's expansion history based on the performance derived in the previous subsection.

The two Universe scale factors at the astrnomical object and the observer are related as follows:
\begin{equation}
	\label{eqn:redshift}
	\frac{a(t_{obs})}{a(t_{obj})} = 1 + z_{obj},	
\end{equation}
where $z_{obj}$ is the redshift of the object, $a$ is the Universe scale factor, and $t_{obj}$ and $t_{obs}$ are the time at the object and the observer, respectively. The change in the redshifts of the source at $z_{obj}$ at two epochs with a time span of $\Delta t_{obj}$ is  
\begin{eqnarray}
	\label{eqn:redshift_drift}
	\Delta z &=& 1 + z_{obj}|_{t=t_{obj} + \Delta t_{obj}} - \left(1 + z_{obj}|_{t=t_{obj}} \right) 	\nonumber \\
	&=& \frac{a(t_{obs}+\Delta t_{obs})}{a(t_{obj}+\Delta t_{obj})} - \frac{a(t_{obs})}{a_{obj}(t_{obj})} \nonumber \\
	&\simeq & \frac{\dot{a}(t_{obs})}{a(t_{obs})} \left(1 + z_{obj} \right) \Delta t_{obs} - \frac{\dot{a}(t_{obj})}{a(t_{obj})} \frac{a(t_{obs})}{a(t_{obj})} \Delta t_{obj},
\end{eqnarray}
where $\dot{a}$ is the time differential of the Universe scale factor. The Taylor expansion was used because the time intervals of $\Delta t_{obj}$ and $\Delta t_{obs}$ are much shorter than $t_{obj}$ and $t_{obs}$. Considering that the comoving distance between the object and the observer is constant, the light emitted from the same astronomical objects at the two time of $t_{obj}$ and $t_{obj}+\Delta t_{obj}$ is observed by the observer at the two time of $t_{obs}$ and $t_{obs}+\Delta t_{obs}$, respectively:
\begin{equation}
	\label{eqn:comoving_distance}
	\int_{t_{obj}}^{t_{obs}} \frac{1}{a(t)} dt  = \int_{t_{obj}+\Delta t_{obj}}^{t_{obs} + \Delta t_{obs}} \frac{1}{a(t)} dt.	
\end{equation}
Here, the time intervals of $\Delta t_{obj}$ and $\Delta t_{obs}$ are very small; hence, $\frac{\Delta t_{obj}}{a(t_{obj})} = \frac{\Delta t_{obs}}{a(t_{obs})}$. The change of the redshift with a time interval of $\Delta t_{obj}$ shown in Equation \ref{eqn:redshift_drift} is 
\begin{eqnarray}
	\label{eqn:redshift_drift_2}
	\Delta z &\simeq & \frac{\dot{a}(t_{obs})}{a(t_{obs})} \left(1 + z_{obj} \right) \Delta t_{obs} - \frac{\dot{a}(t_{obj})}{a(t_{obj})} \Delta t_{obs}, \nonumber \\
	&= & H_{0} (1 + z_{obj}) \Delta t_{obs} - H(z_{obj}) \Delta t_{obs},
\end{eqnarray}
where $H(z)$ is the Hubble parameter, and $H_{0}$ is the Hubble constant at the observer. The Hubble parameter can be directly measured from the redshift drift \citep[e.g.,][]{Liske+2008}:
\begin{equation}
	\label{eqn:hubble_parameter}
	H(z)  = H_{0} (1 + z_{obj}) - \frac{\Delta z}{\Delta t_{obs}}.	
\end{equation}

\begin{figure}
	 \centering
	\includegraphics[scale=0.1,height=7cm,clip]{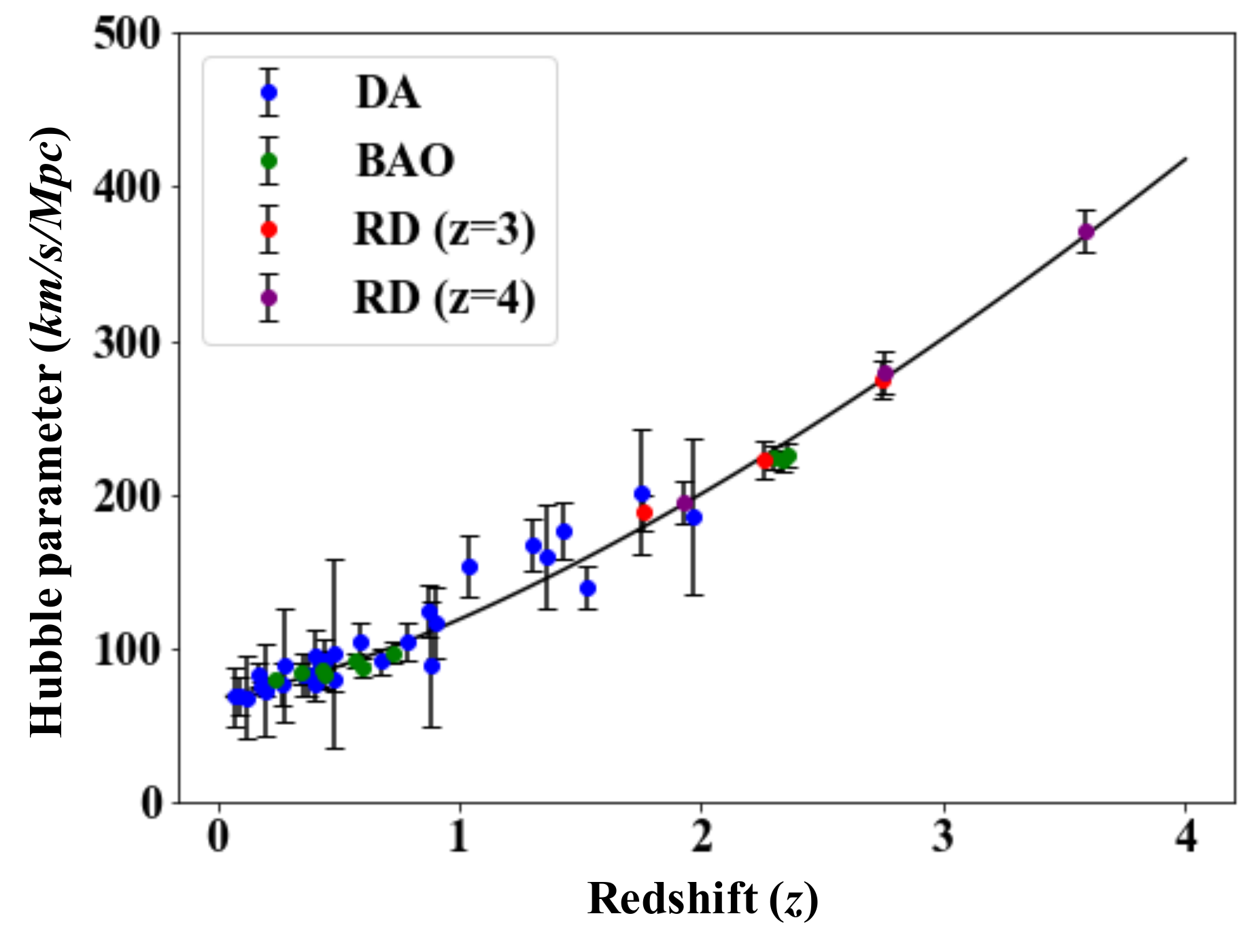}
	\caption{Direct measurement of the Hubble parameter at high redshifts with the proposed instrument concept. The red and purple dots show the redshift drift (RD) measurements for the Lyman alpha forests of the QSOs at  $z=3$ and 4, respectively. The solid line represents $H(z)$ with the parameters of $\Omega_{\lambda}=0.7$, $\Omega_{m} = 0.3$, $H_{0}=67.6km/s/Mpc$. For the comparison, the measurement values from the differential age (DA; red points) and the baryon acoustic oscillation (BAO; blue points) were also plotted \citep{Zhang+2016}. The observing wavelength range was equally divided into 3 for the Lyman alpha forests of the QSOs at $z=3$ and 4. The wavelength intervals for the QSOs at $z$ = 3 and 4 were equal to 603 and 1008 $nm$, respectively. The measurement values were affected by both the shot noise and the systematic noise. The error bar represents the 1-sigma precision limited by the shot noise.}
	\label{fig:hubble_parameter}
\end{figure}

Figure \ref{fig:hubble_parameter} compares the Hubble parameter based on the cosmological model with the redshift drifts derived from the expected performance for the Lyman alpha forests of the QSOs at $z=3$ and 4. Compared to the other methods for measuring the Hubble parameter, such as the differential age (DA) and the BAO, the redshift drifts could constrain the Hubble parameter at high redshifts ($z > 2$). The measurement values derived from the other methods have the uncertainty on the age estimation of galaxies or a cosmological model-dependence. In contrast, this method directly measures the Hubble parameters without any uncertainty and model depndence; hence, the comparison of the measurement values from the other methods with the radshift drift ones is valuable. This direct measurement of the expansion history of the Universe may solve the Hubble tension and discover new physics.

\section{Characterization of the reflected and thermal light from habitable planet candidates} \label{sec:discussion}

We introduce this evaluation method to characterize the reflected and thermal light from the habitable planet candidates orbiting the late-type stars and investigate optimum optical parameters of the spectrograph for this science case. Based on this investigation, we present an optical design of the high-dispersed densified pupil spectrograph. We also discuss the detectability of the reflected light with the proposed instrument concept. 

\subsection{Measurement of the reflected light in visible}

Habitable planet candidates orbiting nearby late-type stars were recently discovered \citep[e.g.,][]{Anglada-Escude+2016, Gillon+2016, Zechmeister+2019}. The reflected light from these planets is embedded in the bright light of their host stars; thus, a high-contrast instrument that spatially resolves the planetary system and mainly suprresses the light of only the host star is required to detect the reflected light. By contrast, because the planetary orbital motion imprints the Doppler shift information on the reflected light, the high-resolution spectroscopy of the reflected light motivates us to estimate the orbital inclination and the true planet mass. A number of the absorption lines in the visible host stellar spectrum could be utilized for this purpose. This approach directly measures the stellar spectrum during the observation; hence, an exact stellar spectrum model is not required. The high-resolution spectrograph could also resolve a number of absorption lines produced by the molecules in the planetary atmosphere. This approach helps us estimate the atmospheric compositions of nearby non-transiting planets. 

We investigated herein the detectability of the reflected light from the nearby temperate Earth-sized planets with a high-precision radial velocity instrument instead of a high-contrast equipment. All of the nearby habitable planet candidates orbit late-type stars; therefore, the target systems were set to the virtual Proxima Centauri and Trappist 1 systems at 5 $pc$ as examples of the middle and late M type stars. The spectra of the host stars were generated through the \texttt{$BT-settle$} model \citep{Allard+2012}. The contrast ratio between the planet and its host star produced for the Proxima Centauri and Trappist 1 systems was used \citep{Lin+2020}. The spectrum of each planet light was derived by multiplying the host stellar spectrum and the contrast ratio. Table \ref{tab:parameters_of_target_system} compiles the target systems parameters. 

Next, we modified the $Q_{photon}$ and $Q_{sys}$ factors for the case where the object light was embedded in the bright background. It was very difficult to spatially resolve the planet and its host star light, even with a very large space telescope, because of the very small angular separations between the nearby late-type stars and their habitable zones. Note that some proposed coronagraphs had very small inner working angles ($\sim 1 \frac{\lambda}{D}$) \citep[e.g.,][]{Guyon+2014, Itoh+2020, Matsuo+2021} and encouraged us to directly detect the reflected light from the habitable planet candidates orbiting nearby late-type stars. The radial velocity signal of the faint reflected light was embedded in the bright stellar light. In this case, the $i$-th pixel signal for the planetary velocity of $V_{p}$ can be approximated as follows:
\begin{equation}
	\label{eqn:signal_ith_element_planet}
	\frac{\delta V_{p}}{c}(i) \approx \frac{N_{s,2}(i)-N_{s,1}(i)}{\lambda(i) \left(\frac{\partial N_{p}}{\partial \lambda} (i) \right)},
\end{equation}
where $N_{s,j}(i)$ is the number of the photons from the host stellar light at $j$-th epoch, and $N_{p}(i)$ is the number of the photons from the reflected light. The uncertainty is rewritten as
\begin{equation}
	\label{eqn:ideal_limit_planet}
	\sigma^{2}_{wav} = \frac{2}{Q_{photon,p}^{2} \sum_{i} N_{s}(i)}.
\end{equation}
The modified $Q_{photon,p}$ factor for this science case is 
\begin{eqnarray}
	\label{eqn:Q_original_planet}
	Q_{photon,p} &=& \frac{\sqrt{\sum_{i} \frac{\lambda^{2}(i)}{N_{s}(i)}\left(\frac{\partial N_{p}}{\partial \lambda}(i)\right)^{2}}}{\sqrt{\sum_{i} N_{s}(i)}} \nonumber \\
	&=& C_{p/s} \frac{\sqrt{\sum_{i} \frac{\lambda^{2}(i)}{f_{s}(i)}\left(\frac{\partial f_{p}}{\partial \lambda}(i)\right)^{2}}}{\sqrt{\sum_{i} f_{s}(i)}},
\end{eqnarray}
where $C_{p/s}$ is defined as the contrast ratio between the maximum values of $N_{p}$ and $N_{s}$ over the wavelength range, and $f_{p}$ and $f_{s}$ are the normalized values by the maximum values. $f_{p}$ and $f_{s}$ are the same order of magnitude; thus, $\frac{\sqrt{\sum_{i} \frac{\lambda^{2}(i)}{f_{s}(i)}\left(\frac{\partial f_{p}}{\partial \lambda}(i)\right)^{2}}}{\sqrt{\sum_{i} f_{s}(i)}}$ was almost equal to the original $Q_{photon}$ factor. Therefore, the uncertainty of the radial velocity measurements was determined by a combination of the contrast ratio, the square root of the number of the host-star photons acquired over the wavelength range, and the original $Q_{photon}$ factor. The $Q_{sys}$ factor for detecting the reflected light embedded in the stellar light can be also rewritten as follows:
\begin{eqnarray}
	\label{eqn:Q'_factor_planet}
	Q_{sys,p} &=& \sqrt{\sum_{i} \left(\frac{\lambda(i)}{N_{s}(i)}\right)^{2} \left(\frac{\partial N_{p}}{\partial \lambda}(i)\right)^{2}} \nonumber \\
	&=& C_{p/s} \sqrt{\sum_{i} \left(\frac{\lambda(i)}{f_{s}(i)}\right)^{2} \left(\frac{\partial f_{p}}{\partial \lambda}(i)\right)^{2}} \nonumber \\
	&\sim& C_{p/s} Q_{sys}.
\end{eqnarray}
As with the $Q_{photon,p}$ factor, the $Q_{sys,p}$ factor was approximately written as a multiplication of the contrast ratio of the planet to its host stellar light, $C_{p/s}$, and the original $Q_{sys}$ factor. 

\begin{figure}
	 \centering
	\includegraphics[scale=0.1,height=7cm,clip]{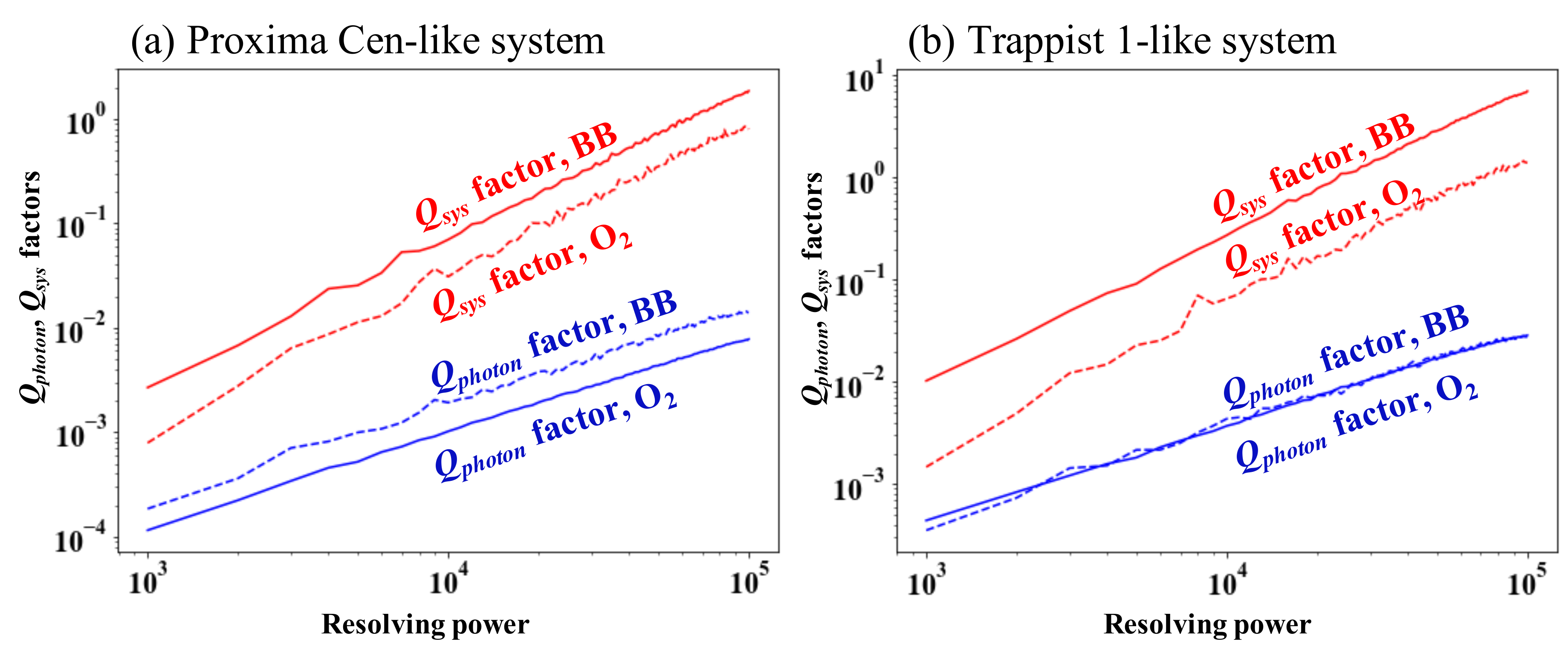}
	\caption{$Q_{photon}$ (blue lines) and $Q_{sys}$ (red lines) factors of the broadband light and the Oxygen A line as a function of the resolving power for the (a) Proxima Cen-like and (b) Trappist 1-like systems. The solid and dashed lines indicate the broadband and Oxygen A line cases, respectively. The broadband light and the Oxygen A line range from 750 to 980 $nm$ and 758 to 770 $nm$, respectively.}
	\label{fig:Q_factors_planet}
\end{figure}

Based on the above consideration, we calculated the $Q_{photon}$ and $Q_{sys}$ factors for each target system, using the planet and its host star spectra simulated by the previous studies. We set the wavelength range to 750 to 980 $nm$ as the broadband light of the reflected light and extracted the wavelength range of 758 to 770 $nm$ to evaluate whether or not the Oxygen A line could be detected as a potential biosignature molecule. Although the Oxygen A line was out of the wavelength range of the instrument concept proposed in Section \ref{sec:evaluation}, the wavelength range can be changed by a small modification of the optical design. Figure \ref{fig:Q_factors_planet} shows the $Q_{photon}$ and $Q_{sys}$ factors as a function of the resolving power. Compared to the original $Q_{photon}$ and $Q_{sys}$ factors shown in Figure \ref{fig:Q_factor}, these factors were reduced by a factor of $10^{6.5}$, almost corresponding to the contrast ratio between the planet and its host star. This result was consistent with what we analytically showed in Equations \ref{eqn:Q_original_planet} and \ref{eqn:Q'_factor_planet}. The $Q_{photon}$ and $Q_{sys}$ factors for the Trappist 1-like system were enhanced because of the more mitigated contrast ratio than the Proxima Cen-like system. 

\begin{figure}
	 \centering
	\includegraphics[scale=0.1,height=7cm,clip]{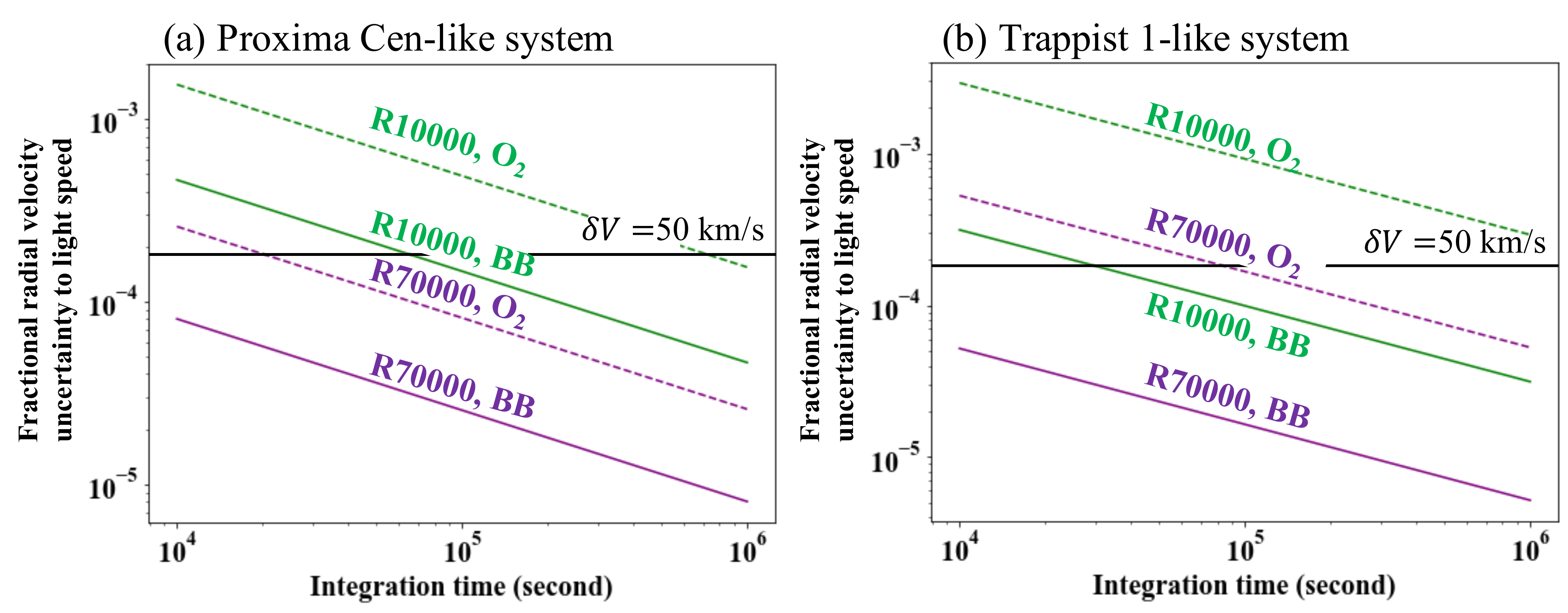}
	\caption{Uncertainty of the radial velocity measurements in two different resolving powers: 10,000 (green lines) and 70,000 (purple lines) as a function of the integration time for the (a) Proxima Cen-like and (b) Trappist 1-like systems. The solid and dashed lines indicate the broadband and Oxygen A line cases, respectively. The horizontal line represents the uncertainty of 50 $km/s$, almost corresponding to the orbital velocities of Proxima Cen b and Trappist 1e. }
	\label{fig:Uncertainty_planet}
\end{figure}

We calculated the uncertainty of the radial velocity measurements from the $Q_{photon}$ factor. Figure \ref{fig:Uncertainty_planet} shows the uncertainty for the broadband light and the Oxygen A line in each target system as a function of the integration time. The orbital velocities of the virtual planets with the same orbital parameters as Proxima Cen b and Trappist 1 e were estimated to be approximately 50 $km/s$ under the parameters compiled in Table \ref{tab:parameters_of_target_system}; thus, the planetary orbital motion and the molecules in the atmosphere can be detected at 1 sigma level when the uncertainty of the radial velocity measurements is less than 0.00017, corresponding to an orbital velocity of 50 $km/s$. Accordingly, 0.8 and 5 days were required as an integration time for detecting the orbital motion and the molecule around the Proxima Cen system with 10,000 resolving power. However, the required integration times could be reduced by a factor of 30 if the resolving power is 7 times higher than that of the instrument concept in Section \ref{sec:concept}. While the $Q_{photon}$ and $Q_{sys}$ factors for the Trappist 1-like system were larger than those for the Proxima Cen-like system, the required integration time was longer due to the limited number of the incident photons. 

\begin{figure}
	 \centering
	\includegraphics[scale=0.1,height=14cm,clip]{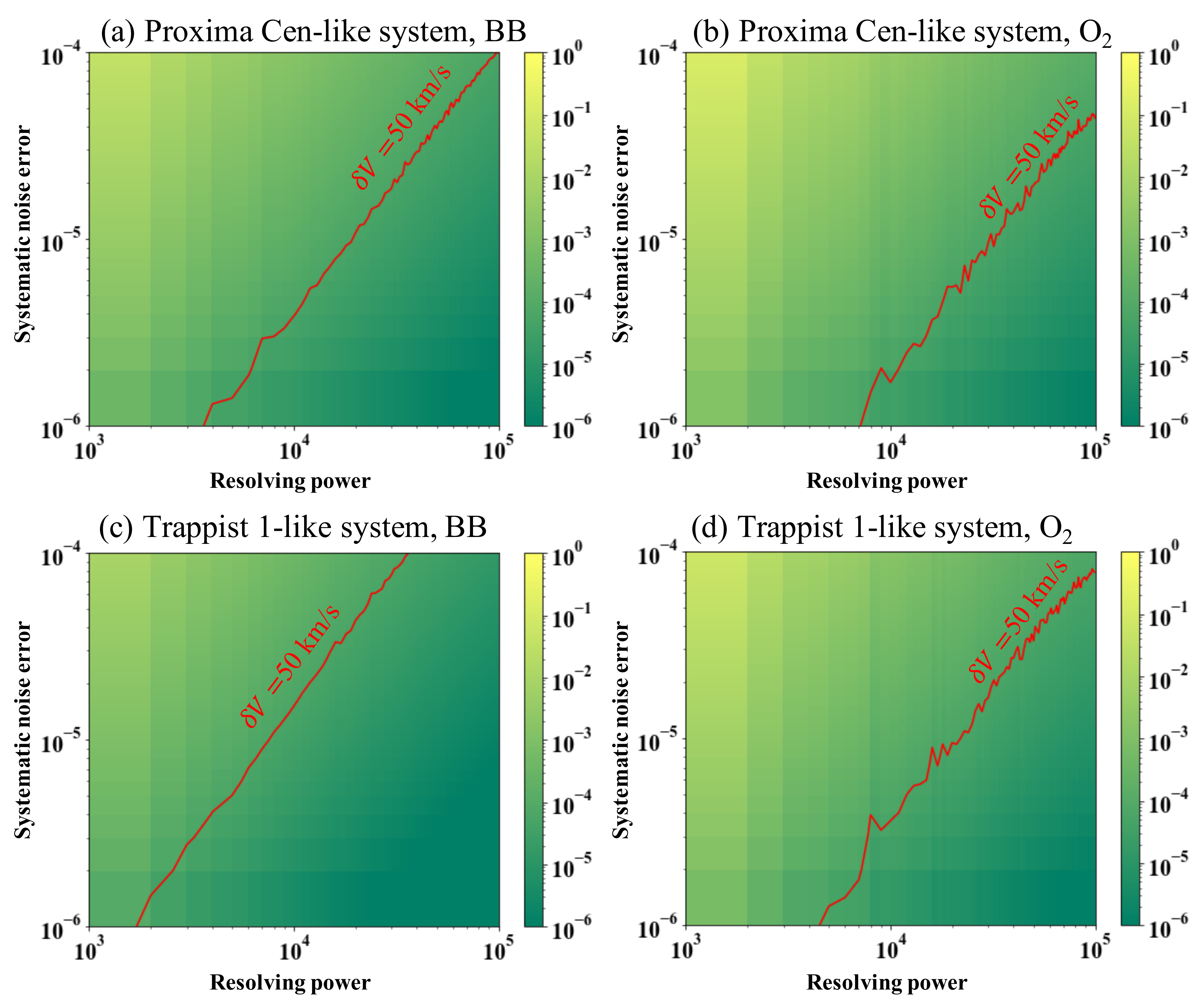}
	\caption{Uncertainty of the radial velocity measurements caused by the systematic noise error in the broadband light and the Oxygen A line for the Proxima Cen-like (upper panels) and Trappist 1-like systems (lower panels) as a function of the resolving power. The red line in each panel represents the 50 $km/s$ uncertainty almost corresponding to the orbital velocities of Proxima Cen b and Trappist 1e.}
	\label{fig:noise_floor_planet}
\end{figure}

Figure \ref{fig:noise_floor_planet} shows the dependency of the noise floor and the resolving power on the uncertainty of the radial velocity measurements. The uncertainty was calculated based on the $Q_{sys}$ factors. The noise floor of a few tens of $ppm$ was acceptable for detecting the planetary orbital motion in each target system under 10,000 resolving power. In contrast, the noise floor should be reduced down to a few $ppm$ to detect the oxygen molecule. The noise floor was also more mitigated for a higher resolving power. When the resolving power is 70,000, the systematic noise floor of 10 $ppm$ level is acceptable. This noise level has been achieved in the room experiments \citep{Matsuo+2019,Schlawin+2021}. In distinction from the science case of the direct measurement of the Universe's expansion history, a resolving power higher than 70,000 is desierable for detecting the planetary orbital motion and the molecules from the reflected light. When the resolving power of this proposed concept is enhanced up to 70,000, this instrument concept could provide a new promising approach for characterizing the reflected light from the non-transiting planets orbiting nearby late-type stars. The observable targets by this approach within a feasible integration time would be middle and late M-type stars within 10 and 5 $pc$, respectively. However, this evaluation ignored the impact of the stellar variability on the photometric precision. How the stellar activity affects the detectability should be investigated as a future work. 

Based on these considerations, the increase in the resolving power of the spectrograph to 70,000 cannot only reduce the required integration time for detecting the atmospheres of exoplanets but also mitigate the condition on the systematic noise error. In the next subsection, we present an optical design of a seven times higher dispersed spectrograph than that for the direct measurement of the Hubble parameters.

The abovementioned results were in the same order of magnitude as those roughly estimated in terms of the signal-to-noise ratio. The signal-to-noise ratio for detecting the planet signal embedded in the stellar light could be improved by the square root of the number of the absorption lines resolved by a high-resolution spectrograph \citep{Snellen+2015}: 
\begin{eqnarray}
	\label{comparison_previous_study}
	\left( \frac{S}{N} \right)_{detection} &\approx & \frac{S_{p}}{\sqrt{S_{s}}}\sqrt{N_{lines}} \nonumber \\
		&=& \frac{C_{p/s}\sqrt{N_{lines}}}{\sqrt{S_{s}}},
\end{eqnarray}
where $S_{p}$ and $S_{s}$ are the numbers of the planet and star photons in each spectral element, respectively, and $N_{lines}$ is the number of absorption lines included in the wavelength range. We assumed that the shot noise caused by the stellar light dominated the detector noise. When the signal-to-noise ratio was equal to 1, the uncertainty of the radial velocity measurements was approximately $\frac{c}{R}$, where $R$ is the resolving power. For example, when the resolving power was 10,000, the number of the resolved lines was approximately 3200 and 100 for the broadband light and the Oxygen A line, respectively. The integration times required for detecting an orbital motion with $\delta V_{p}$ of 30 $km/s$ in the Proxima Cen-like system are 11.4 and 239 h for the broadband light and Oxygen A line cases, respectively. As shown in Figure \ref{fig:Uncertainty_planet}, these values were in a good agreement with those required for reducing the uncertainty of the radial velocity measurements to 0.0001, corresponding to the orbital motion of 30 $km/s$. 

\begin{table}[htb]
	\begin{center}
		\caption{Parameters of the two target systems}
  		\begin{tabular}{| l | c | c |} \hline 
    		& Proxima Centauri-like system & Trappist 1-like system \\ \hline \hline
    		Distance & 5 $pc$ & 5 $pc$ \\ \hline
		    Temperature of host star & 3000 $K$ & 2500 $K$ \\ \hline    
		    Stellar mass & 0.12 $M_{\odot}$ & 0.09 $M_{\odot}$ \\ \hline    		
    		Stellar radius & 0.16 $R_{\odot}$ & 0.12 $R_{\odot}$  \\ \hline
    		Planet radius & 1 $R_{\oplus}$ & 1 $R_{\oplus}$ \\ \hline
    		Semi-major axis & 0.049 $AU$ & 0.029 $AU$ \\ \hline
    		Composition of atmosphere & Earth-like composition  & Earth-like composition \\ \hline
    		Surface pressure of planet & 1 $bar$ & 1 $bar$ \\ \hline
  \end{tabular}
  \label{tab:parameters_of_target_system}
  \end{center}
\end{table}

\subsection{Spectrograph design for charcterization of the atmospheres of habitable planet candidates} \label{subsec:instrument_design_planet}

\begin{figure}
	 \centering
	\includegraphics[scale=0.1,height=8.5cm,clip]{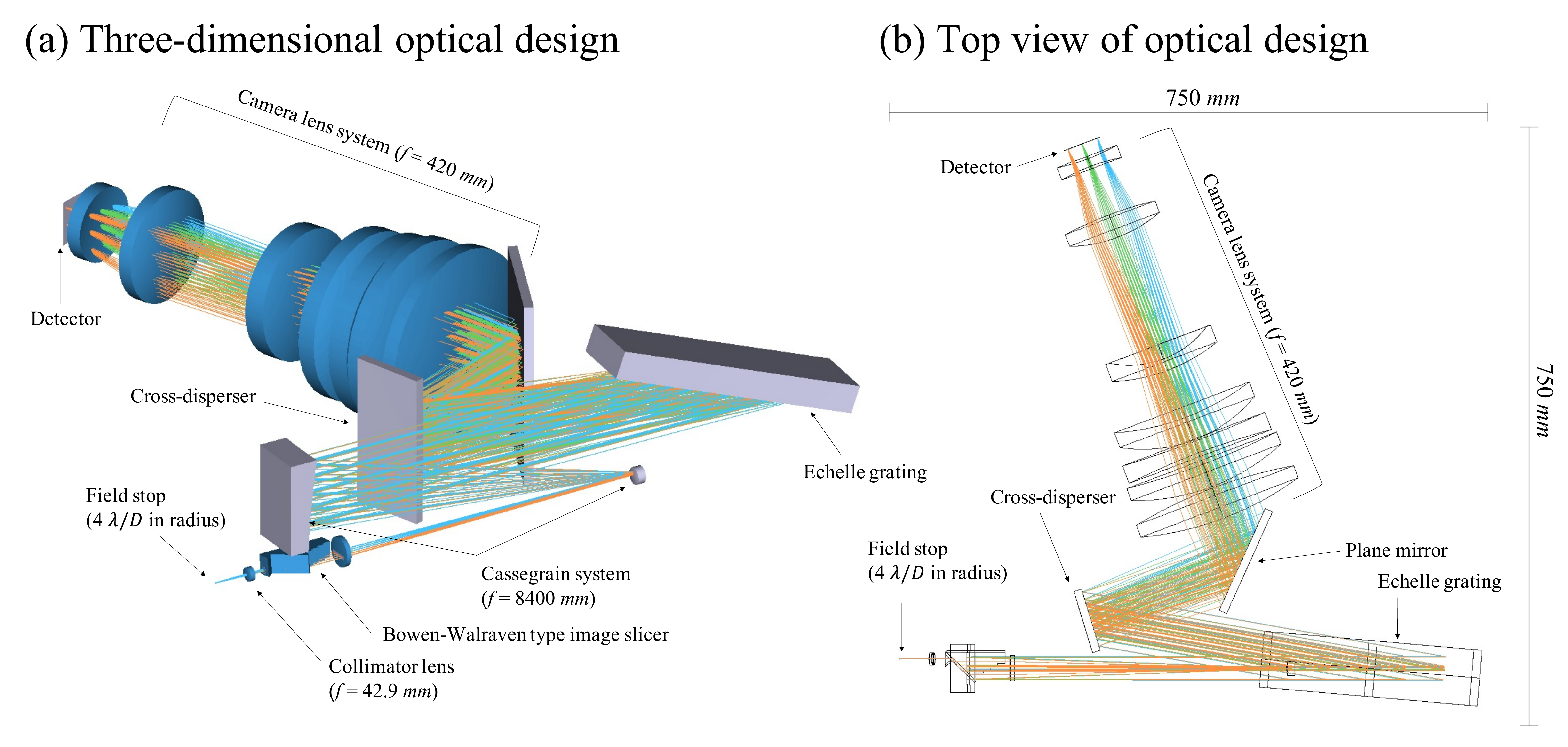}
	\caption{High-dispersion spectrograph design optimized for characterization of the atmospheres of habitable planet candidates. The blue, green, and orange rays indicate the diffraction orders of 83, 72, and 64, respectively. Each ray has a field-of-view of 4 $\frac{\lambda}{D}$ in radius. The entrance beam with 3 $mm$ diameter was divided into five without any densification.}
	\label{fig:High_dispersion_DPS_design_planet}
\end{figure}

We present an optical design of the high-dispersed densified pupil spectrograph for characterizing the atmospheres of habitable planet candidates. The resolving power was set to approximately 70,000, instead of that of 10,000. The wavelength of the spectrograph ranges from 745 to 984 $nm$ to detect the Oxygen A line at 0.76 $\micron$ as a biosignature and a strong water vapor line around 0.95 $\micron$ as an indicator of the habitability. Figure \ref{fig:High_dispersion_DPS_design_planet} shows the optical design of the spectrograph. Although the optical configuration was the same as that for the direct measurement of the Hubble parameters, only the optical parameters were changed to increase the resolving power by a factor of seven. Table \ref{tab:optical_parameters_planet} compiles the parameters of the optical design. 

The volume of the spectrograph was 750 (L) x 750 (W) x 250 (H) $mm$. This small spectrograph design could be utilized for space telescopes. This proposed approach could realize a combination of the high presicion spectrophotometry with the high resolving power, which opens a new path toward characterization of the atmospheres of nearby habitable planet candidates. We discuss the reason why this proposed concept could more reduce the volume of the spectrograph compared to conventional high-dispersion spectrographs with the same field of view in Section \ref{sec:hds_for_space_telescopes}.

The increase in the resolving power of the spectrograph to 100,000 is achievable by optimizing the optical parameters. The required integration time could be more reduced than that shown in Figure \ref{fig:Uncertainty_planet}. However, because the higher resolving power more limits the wavelength range with the same detector format, another spectral channel is required to keep the same wavelength range, which widens the size of the spectrograph. The resolving power should be determined such that the resource of a space telescope and the science merit are balanced.

\begin{figure}
	 \centering
	\includegraphics[scale=0.1,height=12cm,clip]{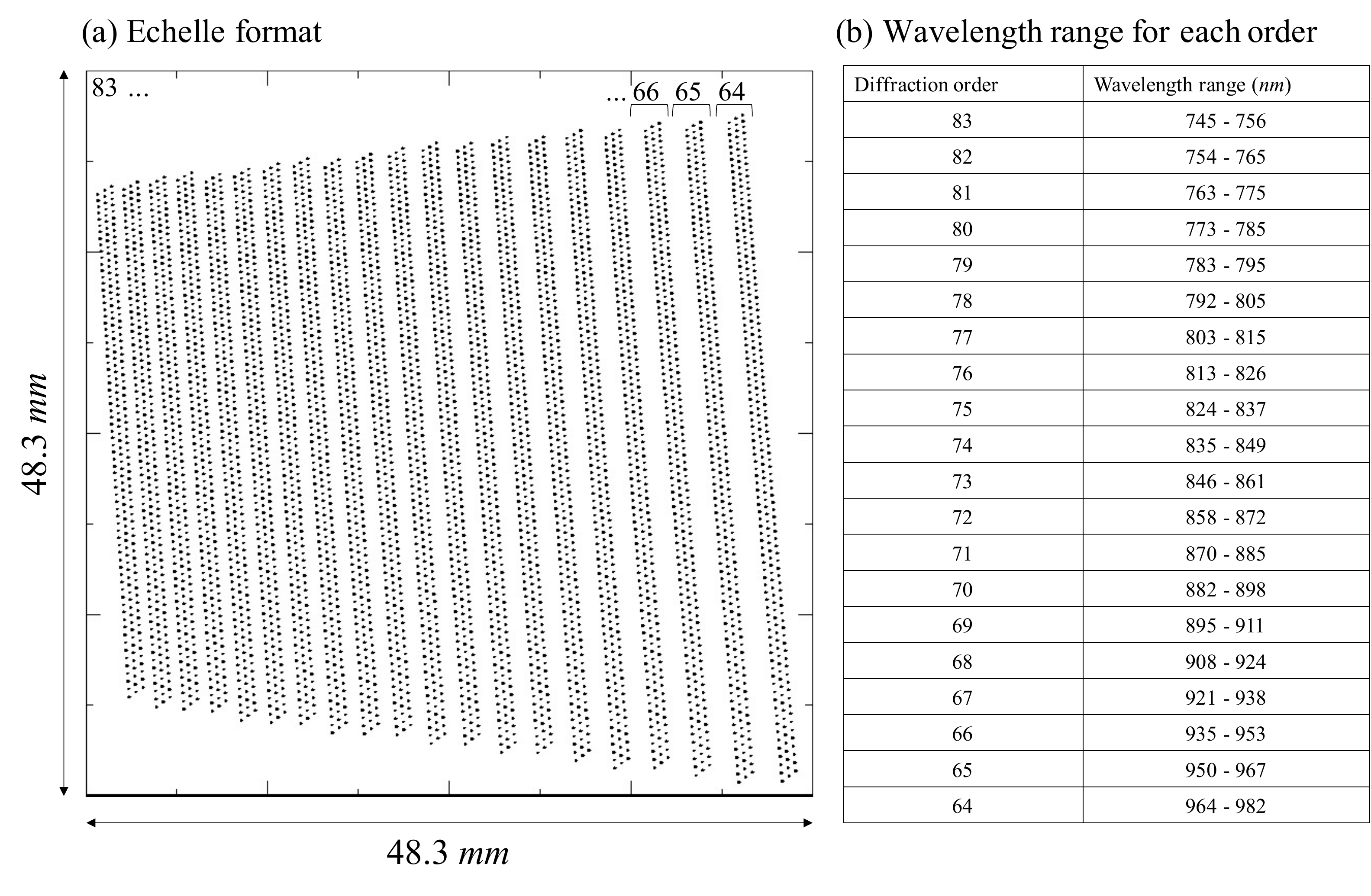}
	\caption{(a) Echelle format on the detector and (b) the wavelength range of each order. The 20 vertical lines show the different diffraction orders of from 64 to 83 indicated by the numbers shown above the lines. The detector size was assumed to be 48.3 x 48.3 $mm$.}
	\label{fig:footprints_planet}
\end{figure}

\begin{table}[htb]
	\begin{center}
		\caption{Parameters of the optical designs for characterization of habitable planet candiates}
  		\begin{tabular}{| l | c |} \hline 
    		Wavelength range & 745 - 982 $nm$ \\ \hline
    		Resolving power at the central wavelength of each order & 71,000 \\ \hline 
    		Field-of-view & 4 $\frac{\lambda}{D}$ at 900 $nm$ \\ \hline     
			Incident beam diameter & 3 $mm$ \\ \hline    		
    		Focal length of the collimator lens & 42.9 $mm$ \\ \hline
    		Number of sub-pupils & 5  \\ \hline
    		Focal length of the Cassegrain system & 8400 $mm$ \\ \hline
    		Type of cross-disperser & Reflection grating \\ \hline
    		Pitch interval of the Echelle grating & 30 $lines/mm$ \\ \hline
    		Brazed angle & 70 $degree$ \\ \hline
    		Diffraction order & 64 - 83 \\ \hline
    		Focal length of the camera lens system & 420 $mm$ \\ \hline
			Number of samplings along spectral direction & 2.5 \\ \hline    		
    		Detector format & 4 k x 4 k \\ \hline
    		Pixel size of the detector & 12 $\micron$ \\ \hline
  \end{tabular}
  \label{tab:optical_parameters_planet}
  \end{center}
\end{table}


\subsection{Measurement of the thermal emissions at mid-infrared wavelengths}

This instrument concept could be also applied to a longer wavelength regime. The densified pupil spectrograph concept was originally proposed for transit spectroscopy at the mid-infrared wavelengths \citep{Matsuo+2016} and applied to the Origins Space Telescope concept that was proposed as one of the four large space mission concepts. This telescope concept provided a very wide wavelength range of approximately 3 to 600 $\micron$, thanks to the telescope and the instruments that cooled down to the cryogenic temperature \citep{Leisawitz+2018}. As one of the baseline instruments, the mid-infrared spectrograph employed the densified pupil spectrograph to search for biosignature molecules through a highly stable transit spectroscopy of Earth-sized planets orbiting late-type stars \citep{Meixner+2019}. Recently, focusing on the fact that the contrast ratio between the temperate planet and its host star is mitigated to $10^{-4}$ beyond 10 $\micron$, \cite{Fujii+2021} showed that the thermal emissions from the non-transiting temperate Earth-sized planets could be detected with a high-dispersion spectrograph mounted on a large cryogenic telescope, such as the Origins Space Telescope. However, the thermal emission is still embedded in the host star in that wavelength regime; therefore, the instrumental systematic noise could obscure the absorption lines formed by the molecules in the planetary atmosphere. Considering that the shot noise caused by the bright host star was in the order of 100 $ppm$ in the numerical simulation in the previous study, the noise floor of the spectrograph should be reduced down to a few tens of $ppm$ to detect the molecules in the atmospheres. This proposed instrument concept could reduce the systematic noise down to a few $ppm$ and increase the resolving power at the same time; thus, this science case could be realized by this instrument concept. 

\section{high-dispersion densified pupil spectrograph for space telescopes}\label{sec:hds_for_space_telescopes}

We discuss why high-dispersion densified pupil spectrograph designs are suitable for space observatories. These designs have two major advantages over general high-dispersion spectrographs that are widely used for the Doppler planet search on ground-based telescopes: 1. Utilization of a high quality PSF provided by a space telescope and 2. Miniaturization of a highly-disperesed densified pupil spectrograph.

\subsection{Utilization of a high quality PSF}

Space telescopes provide stable, high quality PSFs, and the densified pupil spectrograph concept takes advantage of this to minimize light losses and variations due to pointing jitter. We now compare this to modern fiber-fed spectrographs that are traditionally used for precision radial velocity measurements on ground-based telescopes (e.g., ESPRESSO, EXPRES, KPF). These modern spectrographs are typically located in a temperature- and vibration-controlled room, that are fed by a long optical fiber that provides mechanical isolation from the telescope. 

There are two main differences between the two types of spectrographs. One is the amount of light from a target that is largely lost from the observation. Only part of the PSF passes through a signle-mode fiber that feeds a spectrograph, the remaining light is rejected at the edge of the fiber. As shown in Figure \ref{fig:motion_loss}, the photometric variation caused by the motion loss is a few hundred ppm even under a good line-of-sight jitter of 0.05 $\frac{\lambda}{D}$ when the fiber acts as a small field stop on order of the PSF size. In contrast, the field of view (field stop) of the densified pupil spectrograph concept can be larger than 4 $\frac{\lambda}{D}$, depending on the amount of the incident light to the line-of-sight sensor; as the incident light to the sensor is brighter, the field of view could be larger. The larger field of view significantly reduces the photometric variation due to the motion loss as shown in Figure \ref{fig:motion_loss}. 

Another difference is whether the angle of the incident light to the spectrograph is precisely monitored or not. Because the slit viewer mounted on a conventional spectrograph produces images of the slit field on the focal plane (fiber-fed or not), the slit view is not designed to precisely measure the incident angle of the light to the spectrograph. The main role of the slit viewer is to confirm whether the incident light is introduced to the spectrograph. In contrast, the proposed line-of-sight sensor for the densified pupil spectrograph concept can re-form a PSF-like core from the remaining light on the detector. Because the incident angle of the light directly reflects in the position of the core, a positional precision of 0.001 $\frac{\lambda}{D}$ could be achievable, which leads to reduce the systematic error due to the difference of the line-of-sights at two epochs (less movement means less error). Based on these considerations, this proposed densified pupil approach can utilize high-quality PSFs provided by space telescopes to significantly improve photometric precision over modern fiber-fed spectrographs. We note that a multi-mode fiber for wider field of view cannot deliver the high quality PSF to the detector, which will induce an impact on the photometric precision.

\subsection{Miniaturization} \label{subsec:miniaturization}

The size of the spectrograph shown in Section \ref{sec:discussion} (summarized in Table \ref{tab:optical_parameters_planet}) could be fit onto an approximately 750 x 750 $mm$ bench, achieving the resolving power of more than 70,000. We discuss the reason why the volume of the densified pupil spectrograph can be reduced compared to that of conventional high-dispersion spectrographs with the same field of view. As introduced in Section \ref{sec:concept}, the densified pupil spectrograph concept divides a pupil into five slices, and the beam size along the direction orthogonal to the spectral direction can be reduced by the number of the divided subpupils. Although additional small space (80 x 50 $mm$) is needed for a Bowen-Walraven type image slicer working as a pupil slicer, this pupil division allows reducing the size of the spectrograph in two ways. First, the angle between the incident and output beams to the Echelle grating can be smaller. The distance between the Echelle grating and the cross-disperser is inversely proportional to the number of the subpupils; the distance can be reduced by a factor of five compared to conventional high-dispersion spectrographs. Second, when the distance from the Echelle grating to the cross-disperser is reduced, the diameter of the incident beam to the camera lens system can be also reduced by almost the same factor. Thus, this proposed approach can miniaturize the spectrograph, which is useful for space telescopes.

\section{Conclusion} \label{sec:conclusion}

The high-precision radial velocity measurements allowed us to open new science frontiers from the direct measurement of the Universe's expansion history to the characterization of the habitable planet candidates orbiting nearby late-type stars. We introduced herein two important indications, $Q_{photon}$ and $Q_{sys}$, to evaluate how suitable the obtained spectrum is for the radial velocity measurements. While the $Q_{photon}$ factor was used for evaluating the principal limitation on the measurements, the $Q_{sys}$ factor bridged the relation between the systematic noise floor and the precision of the radial velocity measurements. We analytically described how the photometric precision could affect their precision using the $Q_{photon}$ and $Q_{sys}$ factors. The impact of the photometric error on the precision decreased as the resolving power increased. However, the two factors were saturated in the resolving power higher than 10,000 for the Lyman alpha forests of the high-redshift QSOs. The resolving power and the photometric precision limited by the systematic instrumental noise must be simultaneously enhanced.

Based on these considerations, we proposed a new approach to enable high-precision radial velocity measurements over a decade. This instrument concept was composed of a high-dispersion densified pupil spectrograph and a novel line-of-sight sensor minimizing the impact of a telescope pointing error on the spectro-photometric performance. The densified pupil spectrograph produced the spectra of the densified sub-pupils on the detector plane, to which the primary mirror was optically conjugated, and largely reduced the photometric variation caused by the PSF movement on the detector plane due to the telescope line-of-sight error. This was because the position of each spectrum is, in principle, independent of the low-order wavefront errors. 

However, the difference between the line-of-sights at two epochs with a time baseline of a decade still affected the photometric accuracy, even though the densified pupil spectrograph was applied to the radial velocimetry. The light passing through a diffraction grating was aberrated, and speckles were formed on the science detector, on which the densified pupil spectra were recorded. A change in the line-of-sights moved the position of the ASF on the dispersion grating; thus, the speckle pattern changed, leading to the degradation of the photometric precision. Stabilizing the speckle patterns on the detector is very important for the precision improvement. Therefore, the role of the line-of-sight sensor was to accurately measure the telescope line-of-sight and give the measurement value to a tilt compensation device (e.g., a real-time tip-tilt mirror). In contrast, most of the incident light was introduced to the densified pupil spectrograph as the radial velocity measurements to improve the signal-to-noise ratio and reduce the systematic noise caused by the motion loss on the field stop. Therefore, this new line-of-sight sensor precisely measured the line-of-sight from the remaining light without any significant impact on the main lobe of the ASF. In other words, the line-of-sight sensor could perform at the same time when the densified pupil spectrograph is recorded. Note that the remaining light had little information on the line-of-sight; hence, a general slit viewer does not work under the configuration. 

This line-of-sight sensor produces a PSF-like image on its detector from the outer rings of the ASF on the field stop. This can be realized by putting a pupil stop with a diamter that is excatly similar with that of the entrance pupil, which is put after the outer part of the ASF was introduced to the optical sensor system. The PSF-like image was not equal to the original PSF normally formed on the focal plane. The image was formed through a convolution of the Fourier transform of the pupil stop with the ASF formed on the field stop. The PSF-like image position reflected the telescope line-of-sight; therefore, the difference between the line-of-sights at two epochs can be measured. However, the precision of the tilt measurements will be limited by the number of photons available for this sensor type because the remaining light was used. We confirmed that the sensor could fully measure the line-of-sight with 0.001 $\frac{\lambda}{D}$ accuracy for an integration time of a few minutes under an appropriate assumption. We quantitatively derived the impact of the systematic error on the photometric precision based on an analytical wavefront propagation from the diffraction grating to the detector. A photometric precision of 1 $ppm$ is achievable in optical when the RMS of the diffracted wavefront passing through the diffraction grating is a few tens of $nm$. 

We presented herein two compact densified pupil spectrograph designs with resolving powers of 10,000 and 70,000 by advancing the design originally proposed for mid-infrared transit spectroscopy. The former and latter are optimized for the direct measurment of the Universe's expansion history and the characterization of nearby habitable planet candidates, respectively. While the Lyman alpha forest could be fully resolved by the resolving power of 10,000, a seven times higher resolving power is preferable for the characterization of the planet atmospheres to reduce the integration time and mitigate the requirement on the systematic noise floor. 

For the measurement of the Hubble parameters at various redshifts, two channels could cover a wide wavelength range of 305 to 660 $nm$ with an optical throughput of approximately 63 \% (not including the qauntum efficiency and the telescope throughput). Thanks to the line-of-sight monitor, the systematic error due to the difference of the incident angles to the spectrograph at two epochs could be reduced down to 1 $ppm$. Even though the noise floor is limited to the leve of 10 $ppm$ by the detector system, the precision of the radial velocity measurements could be improved to $\sim$ 2 $cm/s$, which is a fundamental limitation caused by the stability of a calibration source. The Hubble parameters could be directly measured at various redshifts ranging of 1.5 to 4 from measurement of a QSO over a decade at the same precision as the indirect methods, such as the baryon acoustic oscillation.

To detect the Oxygen A line as a biosignature and the water vapor lines as an indicator of habitability, we set the wavelength range to 745 - 982 $nm$ and the resolving power to 71,000, which increases $Q_{photon}$. We successfully present a compact optical design of the high-dispersed densified pupil spectrograph, increasing the resolving power by a factor of seven. The volume of the spectrograph could be within 750 (L) x 750 (W) x 250 (H) $mm$. We investigated the possibility of whether or not the reflected light from virtual habitable planet candidates orbiting nearby non-transiting Proxima Cen- and Trappist 1-like stars can be detected with this instrument concept. Assuming that the telescope diameter is 15 $m$, the Oxygen A line could be detected with a feasible integration time for the Proxima Cen- and Trappist 1-like stars within 10 and 5 $pc$, respectively. Because the high resolving power enhances the $Q_{sys}$ factor, the required condition on the systematic noise floor is also mitigated to the level of 10 $ppm$, which was already achieved by a prototype low-dispersion densified pupil spectrograph in an experiment room. Thus, a combination of the high spectral resolving power and the high precision spectrophotometry could open a new path toward characterizing the atmospheres of nearby non-transiting habitable planet candidates without spatially separating the faint planet light from the bright stellar light. 

Furthermore, focusing on the fact that the visible stellar spectrum imprinted on the reflected light has a number of the absorption lines over the wide wavelength range, we could easily collect a sufficient number of photons to detect the reflected light. This approach directly measures the stellar spectrum as well as the reflected light; hence, an exact stellar spectrum model is not required. Consequently, we found that the reflected light, including a number of the stellar absorption lines, could enhance the $Q_{photon}$ and $Q_{sys}$ factors and allow us to measure the planetary orbital motions with only a few hours integration. The instrument design specification was fully satisfied with the detection of the broadband reflected light. Thus, before the characterization of the planet atmospheres is performed, the true planet mass could be determined for the first a few hours. We note that this evaluation ignored the impact of the host stellar activity on the detectability of the planetary orbital motions and atmospheres. A detailed investigation on the detectability of the reflected light with a stable high-dispersion spectrograph is required in the next step.

Finally, we discussed why the highly-dispersed densified pupil spectrograph is suitable for space applications, comparing this concept with conventional fiber-fed high-dispersion spectrographs. This concept utilizes a stable and highly qualified PSF provided by a space telescope. Because the field of view of the spectrograph is more than 4 $\frac{\lambda}{D}$, the qualified PSF could be delivered to the detector without any significant light loss. The modest field of view could reduce the photometric variation due to the motion loss on the focal plane. In addition, the qualified PSF allows us to precisely measure the incident angle of the light to the spectrograph from the remaining light out of the field of view. Furthermore, thanks to the divided sub-pupils, this concept could more reduce the volume of the spectrograph compared to the general high-dispersion spectrographs. This miniaturization will enable to increase the resolving power under a limited resource of the space telescope.





\appendix
\section{Lyman alpha forest spectra}
\label{sec:appendix}

This section briefly describes the simulations and the post-processing steps used to generate the Lyman-$\alpha$ forest spectra used in this proposal. We used the TNG300-1 simulation \footnote{\url{https://www.tng-project.org/}} \citep{Springel:2018, Marinacci:2018, Naiman:2018,Nelson:2018, Pillepich:2018b}. TNG300-1 has a simulation box of $205  h^{-1} Mpc$ (comoving) and contains $2500^3$ dark matter particles and an equal number of gas cells on average.  The dark matter particle mass was $3.98 \times 10^7 h^{-1}M_{\odot}$ and the average target gas cell mass was $7.44 \times 10^6 \ h^{-1}M_{\odot}$. The gravitational softening length, which sets the minimum gravitational force resolution, is $1 \ h^{-1}kpc$. The cosmological parameters were $\Omega_\Lambda = 0.6911$, $\Omega_m =0.3089$, $\Omega_b =0.0486$, $\sigma_8 = 0.8159$, $n_s=0.9667$ and $h=0.6774$, consistent with the Planck 2015 results \citep{Ade:2015}.  The simulation was run using the cosmological simulation code \texttt{AREPO} \citep{Springel:2010} and includes models for gravitational dynamics, hydrodynamics, star formation, supernova winds and black hole accretion and heating. The gravitational dynamics were computed using the TreePM algorithm, while the gas dynmics used a moving mesh on a dynamically constructed Voronoi mesh. The supernova and the black hole wind models were tuned to reproduce several important properties of the galaxy stellar mass function at $z=0$ and reproduced numerous independent observations \citep{Nelson:2018}. For further details on the simulations we refer the reader to \cite{Pillepich:2018a} and \cite{Weinberger:2017}.

TNG300 was post-processed to produce Lyman alpha forest absorption spectra using the \texttt{fake\_spectra} python package\footnote{\url{https://github.com/sbird/fake_spectra}} \citep{Bird:2014a}. The code approximates each gas particle as an individual absorber producing an absorption profile given by the convolution of a Lorentzian and a Gaussian (a Voigt profile). Peculiar velocities and the Hubble flow were included to give the particle the correct displacement in the velocity space. The internal density profile of each particle was assumed to be a spherical top-hat with the same volume as the Voronoi cell. While this marginally distorted the Voronoi mesh on scales of a single cell, we checked that this was not important at $z < 5$. The hydrogen ionization fraction of the gas was computed by assuming the ionization equilibrium and a spatially uniform but time varying ultra-violet background \cite{FaucherGiguere:2009}. Self-shielding was included using the fitting formula of \cite{Rahmati:2013a}.
We generated $1000$ spectra with a pixel resolution of $\sim 1.5$ km/s at redshifts of $3$ and $4$, albeit the availability of the other redshifts. The spectra were generated at random spatial positions in the box. Neither noise nor instrumental broadening was included.

\acknowledgments
We first would like to thank anonymous referee for providing highly valuable comments that improved the manuscript. We also appreciate Dr. Kentaro Nagamine and Dr. Atsushi Nishizawa for helpful comments on the numerical simulation of Lyman alpha forest. We also acknowledge Dr. Satoshi Itoh for useful discussions on the analytical expression on radial velocity method. Finally, we thank Dr. Yuji Ikeda for having comments on immersion gratings for near-infrared astronomy. T.M. is supported by Grand-in-Aid from MEXT of Japan, No. 19H00700.


\bibliography{DPS_RV_20210917}{}
\bibliographystyle{aasjournal}


\end{document}